\documentclass[a4paper,11pt]{article}

\usepackage[T1]{fontenc}
\usepackage{jheppub}
\preprint{USTC-ICTS/PCFT-26-68}

\usepackage{mathtools}
\usepackage{booktabs}
\usepackage{array}
\usepackage{longtable}
\usepackage{microtype}
\usepackage{enumitem}

\allowdisplaybreaks[2]
\setlist[enumerate]{leftmargin=8mm,itemsep=2pt,topsep=3pt}

\usepackage{tikz}
\usetikzlibrary{arrows.meta,decorations.pathmorphing}

\title{Singlet sector of the vector model with angular potential}
\author[a,b]{Srijan Kumar}
\affiliation[a]{ Interdisciplinary Center for Theoretical Study, University of Science and Technology of China, Hefei,
	Anhui 230026, China.\vspace{.1cm} }
\affiliation[b]{Peng Huanwu Center for Fundamental Theory, Hefei, Anhui 230026, China.}

\emailAdd{srijankumar@ustc.edu.cn}

\abstract{We study the singlet sector of large-$N$, $O(N)$ vector model on $S^1_\beta\times S^2_r$ in presence of an angular potential $\Omega$.
We compute the free energy of the model both at the free fixed point as well as the non-trivial fixed point.
The model undergoes a phase transition 
when the temperature $T$ and angular potential attain the relation $T^2r^2\sim N (1-\omega^2)$, where $\omega=r\Omega$ and $r$ is the radius of $S^2$, with $Tr\gg1$, at both the fixed points, with different numerical coefficients. The transition was studied earlier at zero angular potential at $Tr\sim \sqrt{N}$.
The holonomy eigenvalue distribution, having support on the full circle below the transition, develops a gap above it, called the Gross--Witten--Wadia transition.
The angular potential introduces an additional parameter to such a transition, tuning it, the transition can occur even at a fixed temperature. We show that for the free model transition can occur as $1-\omega^2\sim 1/N$ while $Tr\sim O(1)$. For the interacting model the computation is performed at a limit $Tr\gg1$, thus we show the same at a large but fixed temperature which does not scale as $\sqrt{N}$.
In the ungapped phase, the free energy exhibits a double pole near $\omega^2=1$ for both the fixed points.
This is possibly due to the fact that the theory projected onto the
singlets is non-local and therefore need not obey the leading simple pole scaling of a local thermal effective action.
}

\begin{document}
\maketitle

\section{Introduction}
The singlet sector of the $O(N)$ vector model in $3d$ is proposed to admit a holographic dual description in higher spin gravity in $AdS_4$ \cite{Klebanov:2002ja}, followed by a set of works \cite{Leonhardt:2002sn,Petkou:2003zz,Das:2003vw,Sezgin:2003pt,Giombi:2009wh,deMelloKoch:2010wdf,Giombi:2011ya,Giombi:2012ms,Giombi:2013fka,deMelloKoch:2014vnt,deMelloKoch:2014mos,deMelloKoch:2018ivk} strengthening the proposal by providing more evidence and understanding of it. Recently, a derivation to this duality has been provided in \cite{Aharony:2020omh}. On the field-theory side, the singlet constraint is imposed by averaging the thermal partition function over the $O(N)$ holonomy around the Euclidean thermal circle.
Closely related holonomy matrix-model descriptions of large-$N$ thermal singlet sectors and deconfinement/Hagedorn transitions were developed in
\cite{Sundborg:1999ue,Aharony:2003sx,Jain:2013py}.
% Related large-$N$ thermal holonomy matrix models were studied in \cite{Sundborg:1999ue,Aharony:2003sx,Jain:2013py}. \\
\\

In recent interests of thermal conformal field theories(CFT), large $N$ vector models have played a key role by providing tractable interacting CFT examples to test general CFT principles. In this context, the models have so far been studied in the literature without imposing the singlet constraint. The large $N$ simplification of the partition function allows field theory computations of thermodynamic observables. Thermal one point functions of higher spin currents for the $O(N)$ model on $S^1\times R^2$ were computed in \cite{Iliesiu:2018fao}, followed by generalizations to other vector models and with chemical potential \cite{Petkou:2018ynm,David:2023uya,David:2024naf,Kumar:2025txh,Benedetti:2023pbt}. Systematic corrections to the higher spin currents at each order in $\frac1N$ have been developed in \cite{Diatlyk:2023msc}. A thermal bootstrap approach \cite{Barrat:2025wbi} computes thermal OPE coefficients at finite $N$, this is based on the general method developed in \cite{Marchetto:2023fcw,Marchetto:2023xap} for thermal CFT.
A bootstrap method implementing neural networks to study the OPE of the two point function in \cite{Niarchos:2025cdg} and constraining the analytic structure of scalar two point functions using CFT principles in \cite{Barrat:2025nvu} are interesting in a similar context. The model has also been studied on $S^1\times S^2$ 
%The study of the $O(N)$ model on $S^1\times S^2$ in 
\cite{David:2024pir,David:2025tqn}.
% compute the free energy in systematic high and low temperature expansions, showing their equivalence using numerical analysis. Finite size corrections to thermal one point functions are also evaluated. 
\\

For this work, we will focus on the $O(N)$ model with the singlet constraint.
The model on $S^1\times S^2$ \cite{Giombi:2014iua} undergoes the Gross-Witten-Wadia transition \cite{Gross:1980he,Wadia:1980cp} at a temperature $Tr\sim\sqrt{N}$, as shown in \cite{Shenker:2011zf}, where $r$ is the radius of the sphere $S^2$.
Below the transition, the large-$N$ saddle is ungapped and describes the higher spin singlet gas. Above the transition, the holonomy distribution is gapped, and deep in the high-temperature phase the leading thermodynamics approaches that of the unconstrained vector model.
We will consider the model on $S^1\times S^2$ with an angular potential $\Omega$, whose partition function can be given as follows:
\begin{equation}
	Z(\beta,\Omega)=\operatorname{Tr}_{\rm singlet}e^{-\beta(H-\Omega J_z)},\qquad |\Omega r|<1.
\end{equation}
The model is studied using a Hubbard-Stratonovich transformation at large $N$, introducing an auxiliary field in the path integral. A non-zero angular potential breaking $SO(3)$ to $SO(2)$ leads to a polar angle dependent auxiliary field contributing at the leading order in large $N$ \cite{Mauro:2026zus,David:2026yis,Mishra:2026bpi}, unlike the case at $\Omega=0$. The non-zero angular potential can lead to non-trivial modifications of the Gross-Witten-Wadia transition, as observed in the current study.
In the ungapped phase, the free energy of both the free and interacting models admits a double pole as the leading contribution at $\Omega^2r^2=1$. The model with infinite tower of conserved higher spin charges in the singlet projection is non-local \cite{Aharony:2020omh}, the free energy may not follow the thermal effective action \cite{Anand:2025mfh}, which allows only a simple pole at $\Omega^2r^2=1$ as the leading contribution.
We summarize the key results of our work below for both the free and interacting singlet $O(N)$ model.

\subsection*{Summary of the results}

Writing $\omega=\Omega r$, we determine the Gross--Witten--Wadia transition line and the fast-rotation behavior of the ungapped singlet partition function in the free and critical theories.

\begin{figure}[htbp]
\centering
\includegraphics[width=0.82\linewidth]{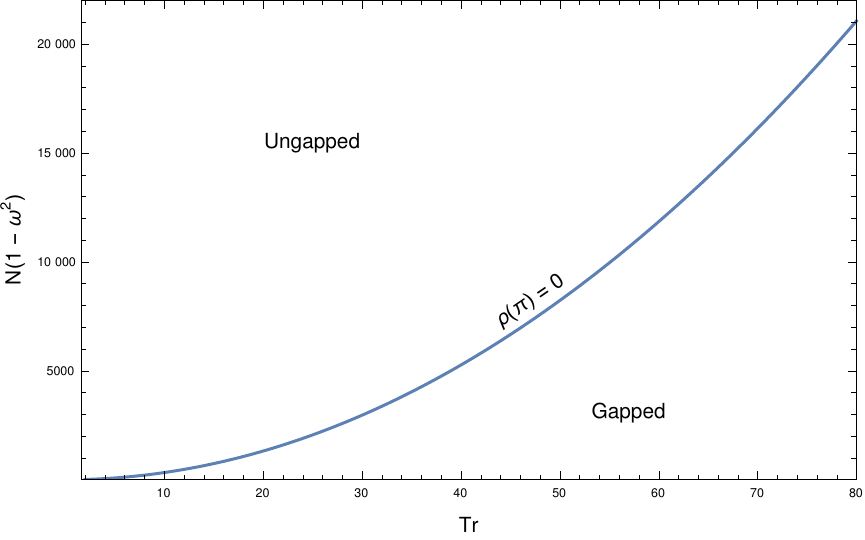}
\caption{Large-$N$ free-theory Gross--Witten--Wadia phase boundary at fixed $Tr$ and $N(1-\omega^2)$. Here $\rho(\alpha)$ is the holonomy eigenvalue density. The curve, defined by $\rho(\pi)=0$, uses the leading term of equation~\eqref{intro-free-fixed-temperature} with the sum truncated at $n=5000$. The regions above and below the curve are ungapped and gapped, respectively.  The figure is used as an illustration of the large $N$ result and should not be interpreted as a finite-$N$ calculation.}
\label{fig:free-double-scaling-transition}
\end{figure}

%For the free theory, the Gross--Witten--Wadia boundary is determined exactly by
%\begin{equation}
%\frac{N}{2}
%=\sum_{\ell=0}^{\infty}\sum_{m=-\ell}^{\ell}
%\frac{1}{\exp\!\left[\frac{\beta}{r}
%\left(\ell+\frac12-\omega m\right)\right]+1}.
%\label{intro-free-exact-transition}
%\end{equation}
For the free theory, the high-temperature transition line is
\begin{equation}
\frac{(Tr)^2}{1-\omega^2}=\frac{3N}{\pi^2},
\qquad
T_c r=\frac{\sqrt{3N(1-\omega^2)}}{\pi}.
\label{intro-free-fixed-omega}
\end{equation}
At fixed $|\omega|<1$, this gives $T_c r=O(\sqrt N)$. At fixed $Tr$, an expansion in small $1-\omega^2$ instead gives
\begin{equation}
1-\omega_c^2
=\frac{2Tr}{N}\sum_{n=1}^{\infty}
\frac{(-1)^{n+1}}{n\sinh[n/(2Tr)]}
+O(N^{-2}).
\label{intro-free-fixed-temperature}
\end{equation}
No high-temperature limit is used to obtain equation~\eqref{intro-free-fixed-temperature}; rotation drives the transition when $1-\omega_c^2=O(N^{-1})$. For $Tr\gg1$, it reduces to $1-\omega_c^2\simeq\pi^2(Tr)^2/(3N)$, consistent with equation~\eqref{intro-free-fixed-omega}. The corresponding phase boundary is shown in figure~\ref{fig:free-double-scaling-transition} as an illustration of the phase transition in the $Tr-\omega$ plane.

Taking $N\to\infty$ first at fixed $Tr$ and $|\omega|<1$, and then $1-\omega^2\to0^+$, the ungapped free partition function develops a double pole:
\begin{equation}
\log Z_{\rm sing}^{N=\infty}
=\frac{(Tr)^2}{2(1-\omega^2)^2}
\sum_{n=1}^{\infty}
\frac{1}{n^3\sinh^2[n/(2Tr)]}
+O\!\left((1-\omega^2)^{-1}\right).
\label{intro-free-double-pole}
\end{equation}
At high temperature, its leading form is
\begin{equation}
\log Z_{\rm sing}^{N=\infty}
=\frac{2\zeta(5)(Tr)^4}{(1-\omega^2)^2}
+O\!\left((1-\omega^2)^{-1}\right).
\label{intro-free-double-pole-high-temperature}
\end{equation}

For the critical theory, the calculation uses the local tangent-plane approximation, valid for $Tr\gg1$. Solving the gap equation together with the condition that the eigenvalue density first vanishes gives
%the rotating saddle is described by $\sqrt{\sigma(\theta)}=Tx\,[1-\omega^2\sin^2\theta]^{-1/2}$. The dimensionless parameter $x$ obeys
%\begin{equation}
%x=\frac{4(Tr)^2}{N(1-\omega^2)}
%\left[
%\operatorname{Li}_3(e^{-2x})
%+x\operatorname{Li}_2(e^{-2x})
%\right]
%-\frac{1}{N}\log\!\left(1-e^{-2x}\right),
%\label{intro-critical-gap}
%\end{equation}
%while the leading smooth-density endpoint condition is
%\begin{equation}
%\frac{1}{2\pi}
%+\frac{2(Tr)^2}{\pi N(1-\omega^2)}
%\left[
%\operatorname{Li}_2(-e^{-x})
%+x\operatorname{Li}_1(-e^{-x})
%\right]=0.
%\label{intro-critical-endpoint}
%\end{equation}
\begin{equation}
\left.\frac{Tr}{\sqrt{N(1-\omega^2)}}\right|_c=0.581068.
\label{intro-critical-numerics}
\end{equation}
%Here $x$ is the dimensionless saddle parameter defined in equation~\eqref{crit-x-definition}. 
At fixed $|\omega|<1$, the transition temperature is $T_c r=0.581068\sqrt{N(1-\omega^2)}$. At fixed $Tr\gg1$, when $Tr$ does not scale as $N$, the critical angular potential satisfies
\begin{equation}
|\omega_c|
=\sqrt{1-\frac{(Tr)^2}{0.337640\,N}},
\qquad
1-\omega_c^2=O(N^{-1}).
\label{intro-critical-fixed-temperature}
\end{equation}
Thus the transition can occur at a temperature that is large in units of $r^{-1}$ but independent of $N$.

In the same ordered limit used for the free theory, the critical saddle at $Tr\gg1$ has the identical leading double-pole coefficient
\begin{equation}
\left.\log Z_{\rm crit,sing}^{N=\infty}\right|_{\rm saddle}
=\frac{2\zeta(5)(Tr)^4}{(1-\omega^2)^2}
+O\!\left((1-\omega^2)^{-1}\right).
\label{intro-critical-double-pole}
\end{equation}
At finite large $N$, the transition occurs before $|\omega|=1$. On the critical-theory transition line, both the double-pole term and the vacuum contribution are $O(N^2)$. The double-pole dominance stated above therefore refers specifically to taking $N\to\infty$ before the fast-rotation limit.

The rest of the paper is organized as follows. Section~\ref{sec:free-model} studies the free singlet model, including the holonomy saddle, the Gross--Witten--Wadia transition, and the free energy. Section~\ref{sec:critical-model} develops the corresponding analysis for the critical model, and section~\ref{sec:discussion} has our discussions. Appendix~\ref{app:haar-measure} derives the large-$N$ orthogonal Haar measure, Appendix~\ref{app:letter-counting} presents the free-theory letter-counting derivation, Appendix~\ref{app:tangent-integrals} evaluates the local tangent-plane integrals, Appendix~\ref{app:tangent-validity} establishes the high-temperature expansion and the validity of the tangent-plane approximation, and Appendix~\ref{app:sigma-determinant} determines the high-temperature scaling of the one-loop determinant.

% FREE SINGLET SECTION LOCKED AFTER SEQUENTIAL REVISION AUDIT (v56)
\section{Singlet sector of the free $O(N)$ model}
\label{sec:free-model}

We begin with $N$ real conformally coupled scalar fields $\phi_i$, transforming in the vector representation of $O(N)$, on $S^1_\beta\times S^2_r$. The action for the model quadratic in $\phi_i$ reads as follows, with $R$ defined as the Ricci scalar of $S^2$.
\begin{equation}\label{free-action}
S_{\rm free}=\frac12\int_{S^1_\beta\times S^2_r}d^3x\,\sqrt{g}\,
\left[g^{\mu\nu}\partial_\mu\phi_i\partial_\nu\phi_i
+\frac{R}{8}\phi_i\phi_i\right],
\qquad R=\frac{2}{r^2}.
\end{equation}
We will evaluate the partition function for this model in presence of the angular potential $\Omega$, in the singlet sector of $O(N)$. The singlet restriction is implemented by averaging over the internal $O(N)$ holonomy, which is described below.

\subsection{Partition function in the singlet sector}

For definiteness, let us consider $N=2K$.\footnote{The odd case $N=2K+1$ gives the same results at leading order in large $N$; its additional one-body Weyl factor is subleading in $N$.} The group $O(2K)$ has two components, with determinants $+1$ and $-1$; the connected component with determinant $+1$ is $SO(2K)$. Although the full singlet projection formally averages over both components, in what follows we evaluate the connected component.\footnote{At fixed operator degree in the strict large-$N$ limit, $SO(N)$ and $O(N)$ singlets coincide; the distinction requires $\epsilon_{i_1\cdots i_N}$. In the ungapped large-$N$ saddle, the $\det U=-1$ component gives the same perturbative holonomy moments and partition function as the connected component. We therefore work with $\det U=+1$.} In this sector, the singlet projection is represented by the $SO(2K)$ Haar integral over $K$ eigenangles $\alpha_a$\footnote{For the $SO(2K)$ eigenangle density, see equation~(3.4) of reference~\cite{Cunden_2018} and section~3.1 of reference~\cite{Meckes_2019}.}.
\begin{equation}
Z_{\rm sing}=\frac{1}{\mathcal N_K}\int_{-\pi}^{\pi}\prod_{a=1}^K d\alpha_a\,
J_{2K}(\{\alpha_a\})\,Z_{\rm matter}(\{\alpha_a\}),
\label{free-singlet-integral}
\end{equation}
where $\mathcal N_K$ is chosen so that the normalized Haar integral of unity equals one. It includes all angle-independent factors and compensates for counting equivalent holonomies more than once when the eigenangles are permuted or an even number of their signs are reversed. The Haar Jacobian is
\begin{equation}
J_{2K}(\{\alpha_a\})=
\prod_{a<b}
\sin^2\frac{\alpha_a-\alpha_b}{2}
\sin^2\frac{\alpha_a+\alpha_b}{2}.
\label{free-haar-jacobian}
\end{equation}

Here $\alpha_a$ are the $K$ eigenangles of the internal holonomy $U\in SO(2K)$ in the fundamental representation. In block-diagonal form, $U$ has eigenvalues
\begin{equation}
\operatorname{spec}(U)=\{e^{\pm i\alpha_1},\ldots,e^{\pm i\alpha_K}\},
\end{equation}
so that
\begin{equation}
\operatorname{Tr}U^n=\sum_{a=1}^K\left(e^{in\alpha_a}+e^{-in\alpha_a}\right)
=2\sum_{a=1}^K\cos(n\alpha_a).
\end{equation}
%We use the eigenangles $\alpha_a$ as integration variables; the Weyl-group overcounting associated with the full integration range is included in $\mathcal N_K$. 
For the vector matter considered below, the holonomy dependence is completely captured by the fundamental traces $\operatorname{Tr}U^n$.

The eigenangles, with $-\pi\leq\alpha_a\leq\pi$, act as imaginary chemical potentials $i\alpha_a/\beta$ for the corresponding internal charges. In this basis, pair the $2K$ real scalars into $K$ complex fields
\begin{equation}
\Phi_a=\frac{\phi_{2a-1}+i\phi_{2a}}{\sqrt2}.
\end{equation}
The fields $\Phi_a$ and $\Phi_a^*$ carry holonomy phases $e^{i\alpha_a}$ and $e^{-i\alpha_a}$, respectively. The single-particle modes of a conformally coupled scalar on $S^2_r$ have energy $E_\ell$ and azimuthal angular momentum $m$, with
\begin{equation}
E_\ell=\frac{\ell+\frac12}{r},\qquad \ell=0,1,2,\ldots,\qquad m=-\ell,\ldots,\ell.
\end{equation}
The zero-point contribution is independent of the holonomy. Keeping this factor explicit, the matter partition function at a fixed holonomy is
\begin{equation}
\begin{aligned}
E_{\rm vac}^{(\Omega)}
&=\frac{N}{2}\sum_{\ell=0}^{\infty}\sum_{m=-\ell}^{\ell}
\left(E_\ell-\Omega m\right)
=\frac{N}{r}\sum_{\ell=0}^{\infty}
\left(\ell+\frac12\right)^2
=\frac{N}{r}\,\zeta\left(-2,\frac12\right)=0,
\\
Z_{\rm matter}(\{\alpha_a\})
&=e^{-\beta E_{\rm vac}^{(\Omega)}}
\prod_{a=1}^K\prod_{\ell=0}^{\infty}\prod_{m=-\ell}^{\ell}
\frac{1}{\left(1-e^{-\beta(E_\ell-\Omega m)}e^{i\alpha_a}\right)
\left(1-e^{-\beta(E_\ell-\Omega m)}e^{-i\alpha_a}\right)}.
\end{aligned}
\end{equation}
Here $\sum_{m=-\ell}^{\ell}m=0$, and the divergent vacuum sum is evaluated by analytic continuation of the Hurwitz zeta function.
Using $-\log(1-x)=\sum_{n\ge1}x^n/n$, define the $n$-dependent single-particle partition sum
\begin{equation}
a_n\equiv\sum_{\ell=0}^{\infty}\sum_{m=-\ell}^{\ell}e^{-n\beta(E_\ell-\Omega m)}.
\end{equation}
Then
\begin{equation}
\log Z_{\rm matter}=\sum_{n=1}^{\infty}\frac{a_n}{n}\operatorname{Tr}U^n
=2\sum_{a=1}^K\sum_{n=1}^{\infty}\frac{a_n}{n}\cos(n\alpha_a),
\end{equation}
and equation~\eqref{free-singlet-integral} becomes
\begin{equation}
Z_{\rm sing}=\frac{1}{\mathcal N_K}\int_{-\pi}^{\pi}\prod_{a=1}^K d\alpha_a\,
J_{2K}(\{\alpha_a\})
\exp\left[2\sum_{a=1}^K\sum_{n=1}^{\infty}\frac{a_n}{n}\cos(n\alpha_a)\right].
\label{free-singlet-matrix-integral}
\end{equation}
The Haar Jacobian depends only on the internal singlet projection, whereas the temperature and angular potential enter through $a_n$.

Using the dimensionless combinations
\begin{equation}
\omega\equiv\Omega r,\qquad \frac{\beta}{r}=\frac{1}{Tr}.
\label{dimensionless-parameters}
\end{equation}
The sum over the magnetic quantum number is
\begin{equation}
\sum_{m=-\ell}^{\ell}e^{n\beta\Omega m}
=\frac{\sinh[(2\ell+1)n\beta\Omega/2]}{\sinh(n\beta\Omega/2)}.
\end{equation}
The remaining geometric sum over $\ell$ yields
\begin{equation}
a_n=
\frac{e^{-n\beta/(2r)}\left(1+e^{-n\beta/r}\right)}
{\left(1-e^{-n(1-\omega)\beta/r}\right)
\left(1-e^{-n(1+\omega)\beta/r}\right)}.
\label{free-an-exact}
\end{equation}

\subsection{Large $N$ eigenvalue density}
\label{sec:free-density}

At finite $N=2K$, the $K$ angles $\alpha_a$ are the independent variables. A given holonomy configuration defines the even empirical density
\begin{equation}
\rho_K(\alpha)=\frac1N\sum_{a=1}^K\left[\delta(\alpha-\alpha_a)+\delta(\alpha+\alpha_a)\right],
\qquad \int_{-\pi}^{\pi}d\alpha\,\rho_K(\alpha)=1.
\end{equation}
Its cosine moments are
\begin{equation}
\rho_n\equiv\int_{-\pi}^{\pi}d\alpha\,\rho_K(\alpha)\cos(n\alpha)
=\frac{2}{N}\sum_{a=1}^K\cos(n\alpha_a)
=\frac{1}{N}\operatorname{Tr}U^n.
\label{free-rho-moments}
\end{equation}
At finite $N$ the infinitely many moments $\rho_n$ are constrained functions of the same $K$ angles. In the large-$N$ limit a smooth saddle is instead described by a continuous even density $\rho(\alpha)$ satisfying
\begin{equation}
\rho(-\alpha)=\rho(\alpha),\qquad \int_{-\pi}^{\pi}d\alpha\,\rho(\alpha)=1,\qquad \rho(\alpha)\ge0,
\end{equation}
with Fourier representation
\begin{equation}
\rho(\alpha)=\frac{1}{2\pi}+\frac1\pi\sum_{n=1}^{\infty}\rho_n\cos(n\alpha).
\label{free-rho-fourier}
\end{equation}
The Haar Jacobian $J_{2K}$ and the density $\rho$ therefore have distinct roles: the former weights holonomy configurations in the group integral, whereas the latter describes a single configuration in the large-$N$ continuum limit. We use the moments $\rho_n$ to rewrite the exponent and parametrize its leading saddle; we do not perform the exact finite-$N$ change of variables from the eigenangles to the moments. For a smooth density, the collective-field measure contributes a term of order $N$ in the exponential~\cite{JevickiSakita1980,DeanMajumdar2008}. It therefore does not affect the leading $O(N^2)$ saddle.

\subsection{Large $N$ saddle of the holonomy integral}

After absorbing the angle-independent term into the normalization $\mathcal N_K$ in equation~\eqref{free-singlet-integral}, the Haar Jacobian can be written as
\begin{equation}
\log J_{2K}
=-\frac12\sum_{n=1}^{\infty}\frac{(\operatorname{Tr}U^n)^2}{n}
+\frac12\sum_{n=1}^{\infty}\frac{\operatorname{Tr}U^{2n}}{n}.
\label{free-haar-traces}
\end{equation}
A derivation using a convergence factor $q^n$, with $0<q<1$, followed by the limit $q\to1^-$ is given in Appendix~\ref{app:haar-measure}, together with the treatment of the removed constant.

Using equation~\eqref{free-rho-moments}, this becomes
\begin{equation}
\log J_{2K}
=-\frac{N^2}{2}\sum_{n=1}^{\infty}\frac{\rho_n^2}{n}
+\frac{N}{2}\sum_{n=1}^{\infty}\frac{\rho_{2n}}{n}.
\end{equation}
Here $\rho_n^2$ is the square of the $n$th moment, whereas $\rho_{2n}$ is the $(2n)$th moment. The first term is the $O(N^2)$ pairwise Haar interaction, and the second is an $O(N)$ one-body contribution specific to the orthogonal group.

The matter determinant contributes
\begin{equation}
\log Z_{\rm matter}=N\sum_{n=1}^{\infty}\frac{a_n\rho_n}{n}.
\end{equation}
Consequently the complete angle-dependent exponent is
\begin{equation}
\log(J_{2K}Z_{\rm matter})
=-\frac{N^2}{2}\sum_{n\ge1}\frac{\rho_n^2}{n}
+N\sum_{n\ge1}\frac{a_n\rho_n}{n}
+\frac{N}{2}\sum_{n\ge1}\frac{\rho_{2n}}{n}.
\label{free-complete-exponent}
\end{equation}
For moments $\rho_n=O(1)$, the pairwise Haar term is $O(N^2)$ and favors the uniform density. The matter term can compete with it only when $a_n=O(N)$. Although the one-body Haar term is then subleading, we retain it in the saddle. Completing the square in equation~\eqref{free-complete-exponent} gives
\begin{equation}
\log(J_{2K}Z_{\rm matter})
=-\frac{N^2}{2}\sum_{n=1}^{\infty}\frac{1}{n}
\left[\rho_n-\frac{1}{N}
\Big(a_n+\frac{1+(-1)^n}{2}\Big)\right]^2
+\frac12\sum_{n=1}^{\infty}\frac{1}{n}
\Big(a_n+\frac{1+(-1)^n}{2}\Big)^2.
\label{free-holonomy-square-completion}
\end{equation}
We now solve for the ungapped density and determine when it first reaches the boundary $\rho(\alpha)=0$. This fixes the transition scale without assuming in advance how the temperature scales with $N$.

For an ungapped saddle with $\rho(\alpha)>0$, positivity does not constrain infinitesimal variations, and the nonzero cosine moments may be varied independently. Extremizing equation~\eqref{free-holonomy-square-completion} with respect to each $\rho_n$ gives
\begin{equation}
0=-\frac{N^2}{n}\rho_n+\frac{N}{n}a_n
+\frac{N}{2n}\left(1+(-1)^n\right).
\end{equation}
Therefore,
\begin{equation}
\rho_n=\frac{a_n}{N}
+\frac{1+(-1)^n}{2N}.
\label{free-rho-saddle}
\end{equation}
If $a_n/N\to0$ for every fixed $n$, these moments vanish and the strict leading saddle is the uniform distribution,
\begin{equation}
\rho_n^{(0)}=0,\qquad \rho^{(0)}(\alpha)=\frac{1}{2\pi}.
\end{equation}
The negative coefficient of the square in equation~\eqref{free-holonomy-square-completion} confirms that the stationary point is a maximum of the exponent. Substitution of equation~\eqref{free-rho-saddle} into the Fourier expansion in equation~\eqref{free-rho-fourier} gives
\begin{equation}
\rho(\alpha)=\frac{1}{2\pi}
+\frac{1}{\pi N}\sum_{n=1}^{\infty}a_n\cos(n\alpha)
+\frac{1}{2\pi N}\sum_{n=1}^{\infty}
\left(1+(-1)^n\right)\cos(n\alpha).
\label{free-rho-density}
\end{equation}
The last term in equation~\eqref{free-rho-density} is the $O(N^{-1})$ one-body Haar correction. Its Fourier series is understood distributionally: it consists of delta functions at $\alpha=0,\pi$ together with an $O(N^{-1})$ constant. In the transition regime the matter term is $O(N^0)$, so this distributional correction does not enter the leading smooth-density endpoint condition.

To determine where the minimum occurs at nonzero rotation, define the Boltzmann factor for the scalar mode $(\ell,m)$ by
\begin{equation}
q_{\ell m}\equiv
\exp\left[-\frac{\beta}{r}\left(\ell+\frac12-\omega m\right)\right],
\qquad \ell=0,1,\ldots,\quad m=-\ell,\ldots,\ell,
\qquad 0<q_{\ell m}<1.
\end{equation}
The last inequality follows from $|\omega|<1$. The corresponding coefficient is
\begin{equation}
a_n
=\sum_{\ell=0}^{\infty}\sum_{m=-\ell}^{\ell}q_{\ell m}^{n}.
\end{equation}
Therefore, the matter-dependent nonconstant part of equation~\eqref{free-rho-density} can be summed as
\begin{equation}
F(\alpha)=\sum_{n=1}^{\infty}a_n\cos(n\alpha)
=\sum_{\ell=0}^{\infty}\sum_{m=-\ell}^{\ell}
\frac{q_{\ell m}\cos\alpha-q_{\ell m}^{2}}
{1-2q_{\ell m}\cos\alpha+q_{\ell m}^{2}}.
\end{equation}
Its derivative is
\begin{equation}
F'(\alpha)
=-\sin\alpha\sum_{\ell=0}^{\infty}\sum_{m=-\ell}^{\ell}
\frac{q_{\ell m}(1-q_{\ell m}^{2})}
{(1-2q_{\ell m}\cos\alpha+q_{\ell m}^{2})^2}<0,
\qquad 0<\alpha<\pi.
\end{equation}
Hence the leading smooth part of $\rho(\alpha)$ decreases monotonically as $\alpha$ runs from $0$ to $\pi$, and its minimum occurs at $\alpha=\pi$. Substituting $\alpha=\pi$ into this smooth part of equation~\eqref{free-rho-density} and using $\cos(n\pi)=(-1)^n$ gives
\begin{equation}
\rho_{\rm smooth}(\pi)
=\frac{1}{2\pi}+\frac{1}{\pi N}\sum_{n=1}^{\infty}(-1)^n a_n
=\frac{1}{2\pi}-\frac{1}{\pi N}
\sum_{\ell=0}^{\infty}\sum_{m=-\ell}^{\ell}
\frac{q_{\ell m}}{1+q_{\ell m}}
\underset{\rm transition}{=}0.
\label{free-endpoint}
\end{equation}
In the second equality we used
\begin{equation}
\sum_{n=1}^{\infty}(-1)^n q_{\ell m}^{n}
=-\frac{q_{\ell m}}{1+q_{\ell m}},
\qquad 0<q_{\ell m}<1.
\end{equation}
Since $\rho_{\rm smooth}(\pi)$ is the global minimum of the leading smooth density, the ungapped saddle is admissible at leading order while $\rho_{\rm smooth}(\pi)\ge0$. The transition occurs when this minimum first touches zero; beyond this point its continuation becomes negative near $\alpha=\pi$ and must be replaced by a gapped eigenvalue distribution.

\subsection{Gross--Witten--Wadia transition}

Multiplying equation~\eqref{free-endpoint} by $\pi N$ gives the phase-boundary equation
\begin{equation}
\frac{N}{2}
=\sum_{n=1}^{\infty}(-1)^{n+1}a_n.
\label{free-exact-phase-boundary}
\end{equation}
This is the exact leading large-$N$ boundary between the ungapped and gapped phases. The alternating sign follows from $\cos(n\pi)=(-1)^n$ and does not imply fermionic matter. The equivalent positive mode representation in equation~\eqref{free-endpoint} shows that the right-hand side is even in $\omega$ and increases with $|\omega|$. If its value at $\omega=0$ is smaller than $N/2$, equation~\eqref{free-exact-phase-boundary} determines a critical value $0<|\omega_{\rm cr}|<1$. Equality at $\omega=0$ means that the transition occurs without rotation, while a larger value means that the ungapped saddle is already inadmissible there.

The exact closed form of $a_n$ was given in equation~\eqref{free-an-exact}. At fixed $|\omega|<1$, its high-temperature limit is
\begin{equation}
a_n=\frac{2(Tr)^2}{n^2(1-\omega^2)}+O(1).
\label{free-an-high-temperature}
\end{equation}
Substituting equation~\eqref{free-an-high-temperature} into equation~\eqref{free-exact-phase-boundary} and using $\sum_{n\ge1}(-1)^{n+1}/n^2=\pi^2/12$ gives
\begin{equation}
\frac{(Tr)^2}{1-\omega^2}=\frac{3N}{\pi^2}.
\label{free-critical-line}
\end{equation}
At fixed $|\omega|<1$, this critical line gives
\begin{equation}
T_c r=\frac{\sqrt{3N(1-\omega^2)}}{\pi},
\end{equation}
whereas at fixed high temperature it gives
\begin{equation}
\omega_{\rm cr}=\Omega_{\rm cr}r
=\sqrt{1-\frac{\pi^2(Tr)^2}{3N}}.
\label{free-omega-critical}
\end{equation}
These are two forms of the same high-temperature critical line. At fixed $|\omega|<1$, the first gives $T_c r=O(\sqrt N)$. Equation~\eqref{free-an-high-temperature} then gives $a_n(T_c)=O(N)$ for every fixed $n$, while the one-body Haar correction retained in equation~\eqref{free-rho-saddle} remains $O(N^0)$ in each mode and is suppressed in the leading endpoint condition. Equation~\eqref{free-omega-critical}, however, is only a high-temperature approximation; the fixed-temperature result follows directly from equation~\eqref{free-exact-phase-boundary}.

At fixed $Tr$, expanding equation~\eqref{free-an-exact} for small $1-\omega^2$ gives
\begin{equation}
a_n
=\frac{Tr}{n(1-\omega^2)\sinh[n/(2Tr)]}+O(1),
\qquad |\omega|\to1^-.
\label{free-an-fast-rotation}
\end{equation}
This expansion applies to either sign of the angular potential. To extract the transition at fixed $Tr$ without assuming $\beta/r\ll1$, return to equation~\eqref{free-exact-phase-boundary}. Its right-hand side can become $O(N)$ only when $|\omega|$ approaches one. Substitution of equation~\eqref{free-an-fast-rotation} gives
\begin{equation}
1-\omega_{\rm cr}^2
=\frac{2Tr}{N}
\sum_{n=1}^{\infty}
\frac{(-1)^{n+1}}{n\sinh[n/(2Tr)]}
+O\!\left(N^{-2}\right),
\qquad
|\Omega_{\rm cr}|=\frac{|\omega_{\rm cr}|}{r}.
\label{free-fixed-temperature-omega-critical}
\end{equation}
Equation~\eqref{free-fixed-temperature-omega-critical} is the leading large-$N$ result at fixed $Tr$. For $Tr\gg1$,
\begin{equation}
\sum_{n=1}^{\infty}
\frac{(-1)^{n+1}}{n\sinh[n/(2Tr)]}
\sim2Tr\sum_{n=1}^{\infty}\frac{(-1)^{n+1}}{n^2}
=\frac{\pi^2}{6}Tr,
\end{equation}
so equation~\eqref{free-fixed-temperature-omega-critical} reduces to
\begin{equation}
1-\omega_{\rm cr}^2
\simeq\frac{\pi^2(Tr)^2}{3N}.
\end{equation}
This agrees with equation~\eqref{free-omega-critical} in their common high-temperature regime. More generally, equation~\eqref{free-fixed-temperature-omega-critical} shows that $1-\omega_{\rm cr}^2=O(1/N)$ at fixed $Tr$. Starting from an ungapped saddle at $\omega=0$, increasing $|\omega|$ drives $\rho_{\rm smooth}(\pi)$ to zero at $|\omega|=|\omega_{\rm cr}|$; beyond this point the saddle is gapped.

At zero angular potential, equation~\eqref{free-critical-line} reduces to
\begin{equation}
T_c r=\frac{\sqrt{3N}}{\pi},
\end{equation}
in agreement with the free-model transition temperature found in \cite{Shenker:2011zf}.

\subsection{Free energy at the saddle}
\label{sec:free-strict-limit}

The $O(N)$ one-body Haar term is subleading for the transition saddle but contributes to the $O(N^0)$ partition function at fixed $Tr$. At the saddle in equation~\eqref{free-rho-saddle}, the square in equation~\eqref{free-holonomy-square-completion} vanishes. Expanding the remaining term produces an $a_n$-independent contribution $\frac14\sum_{n\geq1}1/n$. This constant is generated by completing the square; it is distinct from the angle-independent constant removed in deriving equation~\eqref{free-haar-traces}. Its uncancelled appearance reflects the fact that the collective saddle treatment does not include the exact Jacobian for changing from the finite set of eigenangles to the density variables, nor the Gaussian one-loop determinant of the holonomy fluctuations. The pure-Haar contributions of these omitted factors would complete the normalization of the collective integral. We therefore normalize the mode saddle relative to the matter-free Haar integral, for which $a_n=0$ and the partition function equals one. This cancels the $a_n$-independent saddle contribution, leaving
\begin{equation}
\log Z_{\rm sing}^{N=\infty}
=\frac12\sum_{n=1}^{\infty}\frac{a_n^2+a_{2n}}{n}.
\label{free-strict-partition}
\end{equation}
The combination $(a_n^2+a_{2n})/2$ is the character of the symmetric square of the one-particle scalar Hilbert space. Appendix~\ref{app:letter-counting} gives a direct letter-counting derivation of equation~\eqref{free-strict-partition} and explains separately the origins of the $a_n^2$ and $a_{2n}$ terms. Substituting equation~\eqref{free-an-exact} and simplifying gives
\begin{align}
\log Z_{\rm sing}^{N=\infty}
=\sum_{n=1}^{\infty}\frac1{2n}\Big[&
\frac{\cosh^2[n/(2Tr)]}
{\left(\cosh[n/(Tr)]-\cosh[n\omega/(Tr)]\right)^2}
+\frac{\cosh[n/(Tr)]}
{\cosh[2n/(Tr)]-\cosh[2n\omega/(Tr)]}\Big].
\end{align}
The two terms have different singular behavior in the fast-rotation limit. At fixed $\beta/r$, equation~\eqref{free-an-fast-rotation} shows that the $a_n^2$ term in the strict-$N=\infty$ ungapped partition function gives
\begin{equation}
\log Z_{\rm sing}^{N=\infty}
=\frac{(Tr)^2}{2(1-\omega^2)^2}
\sum_{n=1}^{\infty}\frac{1}{n^3\sinh^2[n/(2Tr)]}
+O\!\left((1-\omega^2)^{-1}\right).
\end{equation}
The coefficient is a convergent sum of positive terms, so the leading double pole cannot cancel among the terms in the sum over $n$. The $a_{2n}$ contribution is only of simple-pole order. Thus the double pole is an exact strict-$N=\infty$ statement at fixed $\beta/r$ within the ungapped phase.

As a check, one can subsequently take the high-temperature limit. Equation~\eqref{free-an-high-temperature} then gives
\begin{equation}
a_n^2=\frac{4(Tr)^4}{n^4(1-\omega^2)^2}+O\!\left((1-\omega^2)^{-1}\right),
\qquad
a_{2n}=O\!\left((1-\omega^2)^{-1}\right),
\end{equation}
and hence
\begin{equation}
\log Z_{\rm sing}^{N=\infty}
=\frac{2\zeta(5)(Tr)^4}{(1-\omega^2)^2}
+O\!\left((1-\omega^2)^{-1}\right).
\label{free-double-pole-high-temperature}
\end{equation}
This is the high-temperature limit of the fixed-$\beta/r$ result above, rather than an independent assumption about the origin of the pole.

The same terms can be evaluated on the high-temperature transition curve. Equation~\eqref{free-critical-line} gives $(Tr)^2/(1-\omega_c^2)=3N/\pi^2$, and equation~\eqref{free-an-high-temperature} therefore yields
\begin{equation}
\begin{aligned}
\left.\frac12\sum_{n=1}^{\infty}\frac{a_n^2}{n}\right|_c
&=\frac{18\zeta(5)}{\pi^4}N^2+O(N),
\\
\left.\frac12\sum_{n=1}^{\infty}\frac{a_{2n}}{n}\right|_c
&=\frac{3\zeta(3)}{4\pi^2}N+O(N^0).
\end{aligned}
\label{free-transition-contributions}
\end{equation}
Thus the transition-scale partition function has the structure
\begin{equation}
\left.\log Z_{\rm sing}\right|_c
=\frac{18\zeta(5)}{\pi^4}N^2+O(N).
\label{free-logZ-transition-scaling}
\end{equation}
The $O(N)$ terms in equation~\eqref{free-logZ-transition-scaling} include the one-body orthogonal-Haar contribution in equation~\eqref{free-transition-contributions} and subleading collective-field and holonomy-fluctuation effects. Hence the $a_n^2$ term is the only $O(N^2)$ saddle contribution at the free-theory transition. This differs from the critical theory, where the Hubbard--Stratonovich vacuum term is also $O(N^2)$. At finite large $N$, the Gross--Witten--Wadia transition occurs before $|\omega|=1$ and cuts off the double-pole divergence.

\section{Singlet sector of the critical $O(N)$ model at large $N$}
\label{sec:critical-model}
\subsection{The model and the large-$N$ variables}
We consider the $O(N) $ vector model for scalars interacting via a quartic coupling term on $S^1\times S^2$. The action for the model is obtained by adding a quartic interaction to the free action in equation~\eqref{free-action}:
\begin{equation}
S=\int d^3x\sqrt g\left[
\frac12(\nabla\phi_i)^2+\frac{R}{16}\phi_i^2+\frac{\lambda}{4N}(\phi_i\phi_i)^2
\right],
\label{crit-microscopic-action}
\end{equation}
$R$ is the Ricci scalar of $S^2$ and given by $R=2/r^2$.

We will study the model in the singlet sector of $O(N)$. The singlet sector can be obtained by the same $O(N)$ group average used in the free theory.
The model admits a pair of conformal fixed points; at $\lambda=0$, i.e. free fixed point, examined in the previous section and a non-trivial fixed point at an infinite coupling. In this section we examine the model at the non-trivial fixed point at large $N$.
Simplification of the path integral for the above action at large $N$ with the use of the Hubbard-Stratonovich transformation allows the study of the model at the infinite coupling.
%  simplifications at large $N$ 
%At large $N$ we introduce a Hubbard--Stratonovich field $\sigma(x)$ to linearize the quartic interaction in equation~\eqref{crit-microscopic-action}:
One can perform the Hubbard-Stratonovich transformation by introducing an auxiliary field $\sigma(x)$ in the path integral of the original action. The quartic interaction term is rewritten using a path integral over the auxiliary field $\sigma$ of an action that is quadratic in the field $\phi$. Thus the path integral of equation~\eqref{crit-microscopic-action} on the field $\phi$ can be rewritten as a path integral of the following action over both $\phi$ and $\sigma$.
\begin{equation}
S_{\rm HS}(\lambda)
=\frac12\int d^3x\sqrt g\,
\phi_i\left(-\nabla^2+\frac{R}{8}+\sigma(x)\right)\phi_i
-\frac{N}{4\lambda}\int d^3x\sqrt g\,\sigma(x)^2,
\label{crit-hs-finite-coupling}
\end{equation}
where the Hubbard--Stratonovich contour is chosen so that integrating out $\sigma$ reproduces the positive quartic interaction in equation~\eqref{crit-microscopic-action}.\\

 We note that at zero angular potential the large $N$ limit allows only the constant mode of $\sigma$ in the partition function. Thus $\sigma$ acts as a mass of the scalars while the position dependent modes of $\sigma$ enter only at subleading orders in large $N$. But we will study the model in presence of a non-zero angular potential where the $SO(3)$ symmetry breaks to $SO(2)$. Therefore it allows the position dependent modes at the leading order in large $N$. Hence we retain the auxiliary field as a function of the position  $ \sigma (x)$.
 
%After the relevant mass deformation is tuned to the Wilson--Fisher fixed point, the universal critical large-$N$ description is obtained in the critical limit in which the coefficient of the local $\sigma^2$ term vanishes. 
%The critical action used below is therefore
%\begin{equation}
%S_{\rm HS}^{\rm crit}
%=\frac12\int d^3x\sqrt g\,
%\phi_i\left(-\nabla^2+\frac{R}{8}+\sigma(x)\right)\phi_i .
%\label{crit-hs-action}
%\end{equation}
At the interacting fixed point we take $\lambda\to\infty$, so the local $\sigma^2$ term in equation~\eqref{crit-hs-finite-coupling} vanishes. Following the same group-average construction as in the free theory, the singlet partition function is
% At the non-trivial fixed point the local $\sigma^2$ term in equation~\eqref{crit-hs-finite-coupling} is absent. For a fixed background $\sigma(x)$, the $N$ scalars are Gaussian, and $\sigma(x)$ enters their quadratic operator as a position-dependent potential. Only at the rotation-invariant saddle is this background constant. At nonzero angular potential, the saddle is instead a spatially varying profile; within each tangent patch its local value plays the role of a local mass squared. The vacuum part of the resulting determinant will be evaluated below in the same local tangent-plane approximation used for the thermal contribution.
%
%The singlet projection and the HS integral must now be performed together:
\begin{equation}\label{crit-singlet-partition}
Z_{\rm crit,sing}(\beta,\Omega)
=\operatorname{Tr}_{\rm singlet}e^{-\beta(H-\Omega L_z)}
=\int d\mu_H(\{\alpha_a\})\int \mathcal D\sigma\,
e^{-S_{\rm eff}[\sigma,\{\alpha_a\};\beta,\Omega]}.
\end{equation}
When $N=2K$, the $N$ real scalar fields can be combined into $K$ complex scalar fields, as in the free theory. The holonomy eigenangle $\alpha_a$ then appears in the determinant for the $a$th complex scalar as an imaginary chemical potential $i\alpha_a/\beta$. Integrating out the fundamental scalars gives
\begin{equation}
S_{\rm eff}[\sigma,\{\alpha_a\};\beta,\Omega]
\equiv\sum_{a=1}^{K}
\operatorname{Tr}_{S^1_\beta\times S^2_r}
\log\left[
-\left(\partial_\tau+\frac{i\alpha_a}{\beta}-\Omega L_z\right)^2
-\nabla_{S^2}^2+\frac{R}{8}+\sigma(x)
\right].
\label{crit-effective-action}
\end{equation}
Next we convert the integral over the $K$ imaginary chemical potentials $\alpha_a$ to an integral over the Fourier modes of the holonomy eigenvalue density and evaluate the partition function in equation~\eqref{crit-singlet-partition} by a simultaneous large-$N$ saddle in the density and the auxiliary field $\sigma$. Before doing so, we need a more suitable form of the effective action in equation~\eqref{crit-effective-action}. The position dependence of $\sigma$ makes the computation more involved than at zero angular potential. In the presence of the angular potential, the remaining $SO(2)$ symmetry allows an axisymmetric leading large-$N$ saddle of the form
% 
%The conjugate holonomy phase is already included in the complex-scalar determinant, so no additional sum over signs is required. The functional trace runs over the spacetime modes, with the critical ultraviolet subtraction understood.
%After integrating out the fundamental scalars, the large-$N$ problem is described by two collective variables,
%\begin{equation}
%\rho(\alpha)\qquad\text{and}\qquad \sigma(x).
%\end{equation}
%The first records the distribution of internal holonomy angles, while the second is the Hubbard--Stratonovich background. Its saddle value determines the spatially dependent scalar gap profile. The large-$N$ solution is a simultaneous saddle in these two variables.
%%
%\subsection{Matter coefficients and the two coupled saddle equations}
%
%In the rotating ensemble, we restrict to static, axisymmetric saddles. The auxiliary field then has the form
\begin{equation}
\sigma=\sigma(\theta).
\label{crit-axisymmetric-sigma}
\end{equation}
Still the exact computation of it remains difficult as it involves angular dependence. Thus we will consider the action in a tangent plane associated with a certain point $(\theta_0,\phi_0)$ on the sphere $S^2$ and treat $\sigma(\theta_0)$ as a constant on the tangent plane. We can solve the saddle point configuration of the auxiliary field $\sigma(\theta_0)$ on the tangent plane. 
The systematic structure the auxiliary field $\sigma(\theta)$ on $S^2$ as an expansion around the point with polar angle $\theta=\theta_0$ can be given as
\begin{align}
	\sigma(\theta)
	&=
	\sigma(\theta_0)
	+(\theta-\theta_0)\sigma'(\theta_0)
	+\frac12(\theta-\theta_0)^2\sigma''(\theta_0)
	+\cdots .
	\label{crit-sigma-local-expansion}
\end{align}

Let
\begin{equation}
	\Gamma(\theta)
	=
	\left[1-\Omega^2r^2\sin^2\theta\right]^{-1/2}.
\end{equation}
Appendix~\ref{app:tangent-validity} derives the leading local
phase-space determinant directly from the exact spherical trace, without
assuming the saddle-point profile for $\sigma(\theta)$.  After solving the
leading saddle below, one may check a posteriori that the dimensionless
combination $\beta^2\sigma/\Gamma^2$ remains finite and that the variation
of the saddle over the local thermal scale is parametrically small.  More
precisely, the resulting saddle satisfies
\begin{equation}
	\frac{\beta^2\sigma(\theta_0)}
	{\Gamma(\theta_0)^2}
	=
	O\!\left((\beta/r)^0\right),
	\qquad
	\frac{\beta^2
		|\sigma(\theta)-\sigma(\theta_0)|
	}{\Gamma(\theta_0)^2}
	=
	O\!\left(\frac{\beta}{r}\right)
	\ll 1 ,
	\label{crit-sigma-leading-scaling}
\end{equation}
over angular separations of the order of the local thermal scale.
The second relation measures the variation in units of the local thermal
scale and remains well defined when $\sigma(\theta_0)=0$; in that case its
left-hand side vanishes on the leading ungapped saddle.
These relations therefore provide an a posteriori consistency check of
the leading local saddle, rather than an assumption used in deriving the
local determinant.
%The condition above further requires this difference to remain small relative to $\beta^2|\sigma(\theta_0)|$ within each tangent-plane patch.
We now discuss the leading saddle of the partition function in equation~\eqref{crit-singlet-partition} in the tangent-plane approximation.
%At latitude $\theta$ the local tangential velocity and the associated kinematic factor are
%\begin{equation}
%v(\theta)=\Omega r\sin\theta,
%\qquad
%\Gamma(\theta)=\frac{1}{\sqrt{1-v(\theta)^2}}
%=\frac{1}{\sqrt{1-\Omega^2r^2\sin^2\theta}}.
%\end{equation}

\subsection*{Local tangent-plane approximation}
The calculation below is performed locally on tangent planes to the sphere. Appendix~\ref{app:tangent-validity} shows that it is the leading term in the high-temperature expansion of the exact spherical determinant. To make the construction explicit, choose an arbitrary latitude $\theta_0$ and introduce local orthonormal coordinates in a small patch,
\begin{equation}
x=r(\theta-\theta_0),
\qquad
y=r\sin\theta_0\,(\phi-\phi_0).
\label{crit-local-coordinates}
\end{equation}
For distances $|x|,|y|\ll r$ the spatial metric is
\begin{equation}
ds_{S^2}^2=dx^2+dy^2+O\!\left(\frac{x}{r},\frac{y}{r}\right),
\label{crit-local-metric}
\end{equation}
Since $\theta-\theta_0=x/r$, the expansion in equation~\eqref{crit-sigma-local-expansion} is controlled by the small parameter $x/r$. Thus, at leading order in the local tangent-plane approximation, the Hubbard--Stratonovich field is frozen at its value at the centre of the patch,
\begin{equation}
\sigma(\theta)\simeq\sigma(\theta_0).
\label{crit-local-sigma}
\end{equation}
The angular potential acts in this patch as a constant boost along the azimuthal tangent direction. With $p_i=-i\partial_i$ and $v(\theta)=\Omega r\sin\theta$, the operator inside the trace-log in equation~\eqref{crit-effective-action} reduces locally as
\begin{align}
&-\Big(\partial_\tau+\frac{i\alpha_a}{\beta}-\Omega L_z\Big)^2
-\nabla_{S^2}^2+\frac{R}{8}+\sigma(\theta)
\longrightarrow
-\Big(\partial_\tau+\frac{i\alpha_a}{\beta}-v(\theta_0)p_y\Big)^2
+p_x^2+p_y^2+\sigma(\theta_0).
\label{crit-local-operator}
\end{align}
Terms suppressed by the curvature scale $r^{-2}$ have been omitted. Equation~\eqref{crit-local-operator} is now the quadratic operator of a complex scalar on $S^1_\beta\times R^2$, with the constant local value $\sigma(\theta_0)$ and imaginary chemical potential $i\alpha_a/\beta$. The combination $v(\theta_0)p_y$ is the additional momentum-dependent contribution to the effective chemical potential induced by rotation. The resulting flat-space determinant can be evaluated directly by performing the Matsubara sum and separating the vacuum contribution from the thermal terms. Denoting the regulated tangent-plane area by $V_2=\int dx\,dy$ and writing
$E_p=\sqrt{p_x^2+p_y^2+\sigma(\theta_0)}$, the result for one complex scalar per unit area is
\begin{align}
\frac{\log Z_a^{(\theta_0)}}{V_2}
&=-\int\frac{d^2p}{(2\pi)^2}\Big[\beta E_p
+\log(1-e^{-\beta[E_p-v(\theta_0)p_y]+i\alpha_a})
+\log(1-e^{-\beta[E_p-v(\theta_0)p_y]-i\alpha_a})\Big],\nonumber\\
&\equiv-\int\frac{d^2p}{(2\pi)^2}\beta E_p+\frac{1}{V_2}\log Z_{a,{\rm th}}^{(\theta_0)}
\label{crit-local-determinant}
\end{align}
The first term in equation~\eqref{crit-local-determinant} is the holonomy-independent vacuum contribution. The other two terms are the thermal contributions of the conjugate holonomy phases $e^{\pm i\alpha_a}$. Expanding them with $-\log(1-x)=\sum_{n\geq1}x^n/n$ gives
\begin{equation}
\frac{1}{V_2}\log Z_{a,{\rm th}}^{(\theta_0)}
=2\sum_{n=1}^{\infty}\frac{\cos(n\alpha_a)}{n}
\int\frac{d^2p}{(2\pi)^2}
e^{-n\beta[E_p-v(\theta_0)p_y]}.
\label{crit-local-thermal-density}
\end{equation}

Equation~\eqref{crit-local-thermal-density} is the thermal
log-partition-function density at the fixed latitude $\theta_0$.  The
passage from this local result to the sphere corresponds to retaining the
leading local phase-space contribution in the high-temperature expansion.
At this order, the background quantities $v(\theta)$ and $\sigma(\theta)$
are evaluated at the centre of each local patch, while curvature and
derivative corrections are omitted.  The discrete spherical mode sum is
therefore replaced locally by the continuum phase-space measure
$V_2\int d^2p/(2\pi)^2$, with the factor $V_2$ cancelling the denominator
in equation~\eqref{crit-local-thermal-density}.  Appendix~\ref{app:tangent-validity}
derives this local phase-space expression directly from the exact spherical
trace and shows that, for fixed $n$ and fixed $|\omega|<1$, the
omitted terms are subleading in the high-temperature expansion.

To obtain the contribution from the whole sphere, we add the local contribution from every patch. A patch $P$, centred at latitude $\theta_P$ and having area $\Delta A_P$, contributes its local density multiplied by $\Delta A_P$. As the patches are taken to be increasingly small, their sum becomes the integral over the sphere:
\begin{align}
&\sum_P\Delta A_P
\left.
\left(\frac{1}{V_2}\log Z_{a,{\rm th}}^{(\theta_0)}\right)
\right|_{\theta_0=\theta_P}
\longrightarrow
\int_{S^2}dA\,
\left.
\left(\frac{1}{V_2}\log Z_{a,{\rm th}}^{(\theta_0)}\right)
\right|_{\theta_0=\theta}
\equiv2\sum_{n=1}^{\infty}\frac{A_n[\sigma]}{n}\cos(n\alpha_a),
\label{crit-patch-assembly}
\end{align}
To evaluate $A_n[\sigma]$ explicitly, use $dA=r^2\sin\theta\,d\theta\,d\phi$. In each patch, $y$ is the local azimuthal coordinate introduced in equation~\eqref{crit-local-coordinates}, and $p_y$ is its conjugate momentum. The azimuthal integral gives $2\pi$. With $d^2p=dp_x\,dp_y$, $p^2=p_x^2+p_y^2$, and $\Gamma(\theta)=[1-v(\theta)^2]^{-1/2}=[1-\Omega^2r^2\sin^2\theta]^{-1/2}$, the coefficient $A_n(\sigma)$ in equation~\eqref{crit-patch-assembly} is given by
\begin{align}
A_n[\sigma]
&\equiv 2\pi r^2\int_0^\pi d\theta\,\sin\theta
\int\frac{d^2p}{(2\pi)^2}
\exp\left[-n\beta\left(\sqrt{p^2+\sigma(\theta)}-v(\theta)p_y\right)\right],
\nonumber\\
&=(Tr)^2\int_0^\pi d\theta\,\sin\theta\,
\exp\left[-\frac{n\sqrt{\sigma(\theta)}}{T\Gamma(\theta)}\right]
\left[
\frac{\Gamma(\theta)^3}{n^2}
+\frac{\sqrt{\sigma(\theta)}}{T}\frac{\Gamma(\theta)^2}{n}
\right].
\label{crit-winding-coefficient}
\end{align}
The momentum integral leading to the second line is evaluated in Appendix~\ref{app:tangent-integrals}. In Appendix~\ref{app:tangent-validity}, equation~\eqref{app:tangent-from-weyl} derives the first line of equation~\eqref{crit-winding-coefficient} from the exact spherical trace, and equation~\eqref{app:exact-local-matching} identifies $A_n[\sigma]$ as its leading high-temperature coefficient. The relative correction is $O((Tr)^{-2})$, as stated in equation~\eqref{app:relative-correction}. Summing equation~\eqref{crit-patch-assembly} over the $K$ complex scalars gives
\begin{equation}
\log Z_{\rm th}[\sigma,\{\alpha_a\}]
=2\sum_{a=1}^{K}\sum_{n=1}^{\infty}
\frac{A_n[\sigma]}{n}\cos(n\alpha_a)
=N\sum_{n=1}^{\infty}\frac{A_n[\sigma]\rho_n}{n}.
\label{crit-thermal-determinant}
\end{equation}
For a vanishing Hubbard--Stratonovich background, $\sigma(\theta)=0$, equation~\eqref{crit-winding-coefficient} reduces to
\begin{equation}
A_n[0]=\frac{2(Tr)^2}{n^2(1-\Omega^2r^2)},
\label{crit-zero-background-winding}
\end{equation}
which reproduces the high-temperature coefficient of the free theory. The corresponding expansion of the exact spherical trace is given in equation~\eqref{app:tangent-free-check}.

The first term in equation~\eqref{crit-local-determinant} gives the holonomy-independent vacuum contribution. Summing it over the $K=N/2$ complex scalars multiplies that momentum integral by $N/2$. In dimensional regularization, $\int d^2p\,\sqrt{p^2+\sigma}/(2\pi)^2=-\sigma^{3/2}/(6\pi)$, and hence
\begin{equation}
\log Z_0[\sigma]
=\frac{N\beta}{12\pi}\int_{S^2}d^2x\sqrt g\,\sigma(x)^{3/2}
=\frac{N\beta r^2}{6}\int_0^\pi d\theta\,\sin\theta\,\sigma(\theta)^{3/2}.
\label{crit-vacuum-contribution}
\end{equation}
Equation~\eqref{app:vacuum-local-leading} in Appendix~\ref{app:tangent-validity}
shows that equation~\eqref{crit-vacuum-contribution} is the leading Weyl
contribution to the exact spherical vacuum term. Equation~\eqref{app:vacuum-relative-correction}
shows that, for $\widehat{\sigma}=\beta^2\sigma=O(1)$, its curvature and
derivative corrections are relatively suppressed by $O((Tr)^{-2})$.

Combining the orthogonal Haar measure, the thermal contribution, and the vacuum contribution, the leading local large-$N$ representation is
\begin{align}
Z_{\rm crit,sing}\simeq\int[D\rho][D\sigma]\,\exp\Bigg\{&
-\frac{N^2}{2}\sum_{n=1}^{\infty}\frac{\rho_n^2}{n}
+\frac{N}{2}\sum_{n=1}^{\infty}\frac{\rho_{2n}}{n}
+N\sum_{n=1}^{\infty}\frac{A_n[\sigma]\rho_n}{n}
\nonumber\\
&\hspace{16mm}
+\frac{N\beta r^2}{6}\int_0^\pi d\theta\,\sin\theta\,\sigma(\theta)^{3/2}
\Bigg\}.
\label{crit-large-N-functional}
\end{align}
The first two terms are the $SO(N)$ Haar contribution already encountered in the free theory. The third term is the thermal scalar determinant expanded as a sum over $n$, and the last term is the holonomy-independent vacuum contribution. 

\subsection{Large $N$ saddle point conditions}
Now we will evaluate the path integral in equation~\eqref{crit-large-N-functional} obtained by combining the leading tangent patch approximation on every point on $S^2$, by finding the saddle point configuration of both $\rho_n$ and $\sigma(\theta)$.
 
%The common moments $\rho_n$ couple the tangent patches, and the saddle is obtained by varying the exponent independently with respect to these moments and the full profile $\sigma(\theta)$.
For fixed $\sigma(\theta)$, variation with respect to $\rho_n$ gives the complete mode-by-mode saddle equation following from the exponent in the above equation~\eqref{crit-large-N-functional},
\begin{equation}
0=-\frac{N^2}{n}\rho_n
+\frac{N}{n}A_n[\sigma]
+\frac{N}{2n}\left(1+(-1)^n\right).
\end{equation}
The last term is the explicit contribution of the one-body orthogonal-Haar term $\frac{N}{2}\sum_{k\ge1}\rho_{2k}/k$; it contributes $N/n$ for even $n$ and vanishes for odd $n$. Therefore, the mode-by-mode saddle relation following from equation~\eqref{crit-large-N-functional} is
\begin{equation}
\rho_n
=\frac{A_n[\sigma]}{N}
+\frac{1+(-1)^n}{2N}. 
\label{crit-holonomy-exact}
\end{equation}
%No scaling of $A_n[\sigma]$ has been assumed in equation~\eqref{crit-holonomy-exact}. In particular, when $A_n[\sigma]=O(N^0)$, both terms on its right-hand side are $O(N^{-1})$ and must be kept at the same order. The explicit one-body term will be omitted only when extracting the leading $O(N^0)$ deformation of the smooth density at the Gross--Witten--Wadia transition.

To compute the saddle point condition by varying the auxiliary field configuration $\sigma(\theta)$ keeping $\rho_n$ fixed with the local tangent patch approximation, we need the following functional derivatives
\begin{equation}
\frac{\delta A_n}{\delta\sigma(\theta)}
=-\frac{r^2}{2}\sin\theta\,\Gamma(\theta)
\exp\left[-\frac{n\sqrt{\sigma(\theta)}}{T\Gamma(\theta)}\right].
\end{equation}
Within this leading local approximation, the derivative of the vacuum contribution is
\begin{equation}
\frac{\delta}{\delta\sigma(\theta)}
\left[
\frac{N\beta r^2}{6}\int_0^\pi d\theta'\,\sin\theta'\,\sigma(\theta')^{3/2}
\right]
=\frac{N\beta r^2}{4}\sin\theta\sqrt{\sigma(\theta)}.
\end{equation}
Thus varying the leading local exponent given in equation~\eqref{crit-large-N-functional} at fixed $\rho_n$ therefore gives the leading functional gap equation
%\begin{equation}
%\frac{\sqrt{\sigma(\theta)}}{T}
%=2\Gamma(\theta)\sum_{n=1}^{\infty}\frac{\rho_n}{n}
%\exp\left[-\frac{n\sqrt{\sigma(\theta)}}{T\Gamma(\theta)}\right].
%\end{equation}
%Dividing the gap equation by $\Gamma(\theta)$ gives
\begin{equation}
\frac{\sqrt{\sigma(\theta)}}{T\Gamma(\theta)}
=2\sum_{n=1}^{\infty}\frac{\rho_n}{n}
\exp\left[-\frac{n\sqrt{\sigma(\theta)}}{T\Gamma(\theta)}\right].
\label{crit-gap-redshifted}
\end{equation}
For $\rho_n\geq0$, the left-hand side of equation~\eqref{crit-gap-redshifted} is an increasing function of $\sqrt{\sigma(\theta)}/[T\Gamma(\theta)]$, whereas the right-hand side is decreasing. Its non-negative solution is therefore unique and independent of latitude. We denote this dimensionless constant by
\begin{equation}
x\equiv\frac{\sqrt{\sigma(\theta)}}{T\Gamma(\theta)}.
\label{crit-x-definition}
\end{equation}
Equation~\eqref{crit-gap-redshifted} then reduces to
\begin{equation}
x
=2\sum_{n=1}^{\infty}\frac{\rho_n}{n}
e^{-nx}.
\label{crit-planar-gap}
\end{equation}
%At $\Omega=0$, where $\Gamma=1$, $Tx$ is the uniform solution of the planar gap equation \cite{Shenker:2011zf}.
 More generally, equation~\eqref{crit-x-definition} gives $\sqrt{\sigma(\theta)}=Tx\Gamma(\theta)$. Since $\Gamma(0)=1$, one has $Tx=\sqrt{\sigma(0)}$, and the leading local saddle profile is
\begin{equation}
\sigma(\theta)=\sigma(0)\Gamma(\theta)^2
=\frac{\sigma(0)}{1-\Omega^2r^2\sin^2\theta}.
\label{crit-sigma-profile}
\end{equation}
Equation~\eqref{app:tangent-saddle-consistency} in Appendix~\ref{app:tangent-validity} shows that the dimensionless variation of this profile over a local thermal distance is $O(\beta/r)$. This verifies the local-constancy condition in equation~\eqref{crit-sigma-leading-scaling}.
With $\omega=\Omega r$, substituting $\sqrt{\sigma(\theta)}=Tx\Gamma(\theta)$ into the definition of $A_n[\sigma]$ in equation~\eqref{crit-winding-coefficient} gives
\begin{align}
A_n
&=(Tr)^2\frac{1+nx}{n^2}e^{-nx}
\int_0^\pi d\theta\,\sin\theta\,\Gamma(\theta)^3
\nonumber\\
&=\frac{2(Tr)^2}{1-\omega^2}
\frac{1+nx}{n^2}e^{-nx},
\label{crit-An-onshell}
\end{align}
where
\begin{equation}
\int_0^\pi d\theta\,\sin\theta\,\Gamma(\theta)^3
=\frac{2}{1-\omega^2}.
\end{equation}
Combining the exact holonomy saddle in equation~\eqref{crit-holonomy-exact} with equations~\eqref{crit-planar-gap} and~\eqref{crit-An-onshell} gives
\begin{align}
x
={}&\frac{4(Tr)^2}{N(1-\omega^2)}
\sum_{n=1}^{\infty}\frac{1+nx}{n^3}e^{-2nx}
+\frac{1}{N}\sum_{n=1}^{\infty}
\frac{1+(-1)^n}{n}e^{-nx}
\nonumber\\
={}&\frac{4(Tr)^2}{N(1-\omega^2)}
\left[
\operatorname{Li}_3\left(e^{-2x}\right)
+x\operatorname{Li}_2\left(e^{-2x}\right)
\right]
-\frac{1}{N}\log\!\left(1-e^{-2x}\right).
\label{crit-gap-reduced}
\end{align}
Equation~\eqref{crit-gap-reduced} coincides with the nonrotating gap equation examined in reference~\cite{Shenker:2011zf} under the replacement
\begin{equation}
	\frac{Tr}{\sqrt N}\longrightarrow
	\frac{Tr}{\sqrt{N(1-\omega^2)}}.
	\label{crit-rotation-replacement}
\end{equation}

If $1\ll(Tr)^2/(1-\omega^2)\ll N$, the solution of equation~\eqref{crit-gap-reduced} satisfies $x\to0$ at large $N$.\footnote{When $(Tr)^2/(1-\omega^2)=O(N^0)$, the small-$x$ solution of equation~\eqref{crit-gap-reduced} is $x=[\log N-\log\log N+O(1)]/N$, and therefore $x=O((\log N)/N)$.

} In this regime the leading large-$N$, high-temperature sector is indistinguishable from that of the free singlet theory discussed in the previous section. When $(Tr)^2/(1-\omega^2)=O(N)$, equation~\eqref{crit-gap-reduced} instead admits $x=O(N^0)$; this scaling is considered in the following subsection.
%For the ungapped saddle used in the ordered fast-rotation limit, we first take $N\to\infty$ while keeping $(Tr)^2/(1-\omega^2)$ finite. For any nonzero $N$-independent value of $x$, the right-hand side of equation~\eqref{crit-gap-reduced} then vanishes as $N^{-1}$. The limiting solution is therefore
%\begin{equation}
%\lim_{\substack{N\to\infty\\ (Tr)^2/(1-\omega^2)\ \mathrm{finite}}}x=0.
%\label{crit-x-ungapped}
%\end{equation}
%The logarithmic term in equation~\eqref{crit-gap-reduced} controls only the finite-$N$ approach to this solution.
%

\subsection{Gross-Witten-Wadia transition}
\label{sec:critical-line}
We now use the saddle point conditions derived above to locate the transition. Substituting equations~\eqref{crit-holonomy-exact} and~\eqref{crit-An-onshell} into the Fourier expansion of the eigenvalue density in equation~\eqref{free-rho-fourier} gives
\begin{align}
\rho(\alpha)={}&\frac{1}{2\pi}
+\frac{2(Tr)^2}{\pi N(1-\omega^2)}
\sum_{n=1}^{\infty}
\frac{1+nx}{n^2}e^{-nx}\cos(n\alpha)
+\frac{1}{2\pi N}\sum_{n=1}^{\infty}
\left(1+(-1)^n\right)\cos(n\alpha)
%\nonumber\\
%={}&\frac{1}{2\pi}
%+\frac{(Tr)^2}{\pi N(1-\omega^2)}
%\Big[
%\operatorname{Li}_2\!\left(e^{-x+i\alpha}\right)
%+\operatorname{Li}_2\!\left(e^{-x-i\alpha}\right)
%\nonumber\\
%&\hspace{34mm}
%-x\log\!\left(1-2e^{-x}\cos\alpha+e^{-2x}\right)
%\Big]
%-\frac{1}{2\pi N}.
\label{crit-rho-density}
\end{align}
The value of $x$ is fixed by the full gap equation~\eqref{crit-gap-reduced}. At the transition, the second term in equation~\eqref{crit-rho-density} must cancel the $O(N^0)$ uniform term. This requires $(Tr)^2/[N(1-\omega^2)]=O(N^0)$ and, by equation~\eqref{crit-gap-reduced}, implies $x=O(N^0)$.
The third term in equation~\eqref{crit-rho-density} is suppressed relative to the first two terms by $1/N$
\footnote{The third term in equation~\eqref{crit-rho-density} can be evaluated as
	\begin{equation}
	\frac{1}{\pi N}\sum_{n=1}^{\infty}\cos(2n\alpha)
	=\frac{1}{2N}\sum_{k\in\mathbb Z}\delta(\alpha-k\pi)
	-\frac{1}{2\pi N}.
	\label{crit-distributional-haar-term}
	\end{equation}
	At $\alpha=0,\pi,\ldots$, the delta function is singular, but $\rho(\alpha)$ is understood distributionally. After integration over a range of $\alpha$, the $1/N$ suppression relative to the first two terms in equation~\eqref{crit-rho-density} is manifest.}, and it contains no other parameter that can make it compete with the other two terms.
	It remains to determine where the leading smooth density is minimized. At the transition $x>0$, so define $q=e^{-x}\in(0,1)$. The derivative of the nonconstant smooth term in equation~\eqref{crit-rho-density} is proportional to
$-\sum_{n\geq1}q^n\sin(n\alpha)/n-x\sum_{n\geq1}q^n\sin(n\alpha)$. For $0<\alpha<\pi$, both sums are positive because
\begin{align}
\sum_{n=1}^{\infty}\frac{q^n}{n}\sin(n\alpha)
&=\arctan\!\left(\frac{q\sin\alpha}{1-q\cos\alpha}\right)>0,
\nonumber\\
\sum_{n=1}^{\infty}q^n\sin(n\alpha)
&=\frac{q\sin\alpha}{1-2q\cos\alpha+q^2}>0.
\label{crit-monotonicity-sums}
\end{align}
Consequently, the leading smooth density decreases monotonically from $\alpha=0$ to $\alpha=\pi$ and has its minimum at $\alpha=\pi$. The Gross--Witten--Wadia transition occurs when this minimum first vanishes:
\begin{equation}
\frac{1}{2\pi}
+\frac{2(Tr)^2}{\pi N(1-\omega^2)}
\left[
\operatorname{Li}_2(-e^{-x})
+x\operatorname{Li}_1(-e^{-x})
\right]=0.
\label{crit-endpoint}
\end{equation}
At the leading order in the large-$N$ transition regime, the endpoint condition likewise coincides with its nonrotating counterpart under the replacement in equation~\eqref{crit-rotation-replacement}. 
%Equations~\eqref{crit-gap-reduced} and~\eqref{crit-endpoint} retain the logarithmic correction and determine its finite-$N$ effect. 
Solving equation~\eqref{crit-endpoint} simultaneously with the leading large-$N$ limit of equation~\eqref{crit-gap-reduced}, with $(Tr)^2/[N(1-\omega^2)]=O(N^0)$, gives
\begin{equation}
\begin{aligned}
\left.\frac{Tr}{\sqrt{N(1-\omega^2)}}\right|_c
&=0.581068,
\qquad
x_c=\left.\frac{\sqrt{\sigma(0)}}{T}\right|_c= 0.644715,\\
\left.\frac{r^2\sigma(0)}{N(1-\omega^2)}\right|_c
&=0.140342.
\end{aligned}
\label{crit-numerics}
\end{equation}

The first relation in equation~\eqref{crit-numerics} determines the critical curve. At fixed $\omega$, the transition temperature is
\begin{equation}
T_c r=0.581068\sqrt{N(1-\omega^2)}.
\label{crit-Tc}
\end{equation}
At $\omega=0$, this reduces to
\begin{equation}
T_c(0)r=0.581068\sqrt N,
\end{equation}
whereas at a fixed high temperature and fixed radius, such that $1\ll Tr\ll0.581068\sqrt N$, the critical angular potential is
\begin{equation}
\omega_c=\Omega_c r
=\sqrt{1-\frac{(Tr)^2}{(0.581068)^2N}}
=\sqrt{1-\frac{(Tr)^2}{0.337640\,N}}.
\label{crit-omegac}
\end{equation}
For $Tr\geq0.581068\sqrt N$, the nonrotating saddle is already gapped. For $Tr<0.581068\sqrt N$, it is ungapped at $\omega=0$, and the Gross--Witten--Wadia line is reached by increasing $|\omega|$ to the value in equation~\eqref{crit-omegac}. Equation~\eqref{app:critical-relation-corrections} shows directly that the corrections to this critical relation remain controlled in the correlated high-temperature and fast-rotation limit. The transition calculation is therefore controlled for $Tr\gg1$.

The final relation in equation~\eqref{crit-numerics}, together with the saddle profile in equation~\eqref{crit-sigma-profile}, gives
\begin{equation}
\sigma_c(\theta)=\frac{0.140342\,N}{r^2}
\frac{1-\omega^2}{1-\omega^2\sin^2\theta}.
\label{crit-profile-critical}
\end{equation}
For comparison, the free-theory critical line in equation~\eqref{free-critical-line} gives
\begin{equation}
T_c^{\rm free}r=\frac{\sqrt3}{\pi}\sqrt{N(1-\omega^2)}
\simeq0.551329\sqrt{N(1-\omega^2)}.
\end{equation}

\subsection{Free energy at the saddle}
\label{sec:critical-free-energy}
We now evaluate the logarithm of the partition function, and hence the free energy, at the ungapped saddle. The part of the exponent in equation~\eqref{crit-large-N-functional} that depends on $\rho_n$ can be written as
\begin{equation}
\begin{aligned}
&-\frac{N^2}{2}\sum_{n=1}^{\infty}\frac{\rho_n^2}{n}
+N\sum_{n=1}^{\infty}\frac{\rho_n}{n}
\left[A_n+\frac{1+(-1)^n}{2}\right]
\\
&\qquad={}-\frac{N^2}{2}\sum_{n=1}^{\infty}\frac{1}{n}
\left[\rho_n-\frac{1}{N}\left(A_n+\frac{1+(-1)^n}{2}\right)\right]^2
+\frac12\sum_{n=1}^{\infty}\frac{1}{n}
\left[A_n+\frac{1+(-1)^n}{2}\right]^2.
\end{aligned}
\label{crit-holonomy-square-completion}
\end{equation}
The first term in equation~\eqref{crit-holonomy-square-completion} vanishes at the mode saddle in equation~\eqref{crit-holonomy-exact}. Expanding the remaining term gives the cross term $\frac12\sum_{n\geq1}A_{2n}/n$ and a matter-independent pure-Haar term. The latter is treated as in the free calculation leading to equation~\eqref{free-strict-partition}. Because the collective treatment does not evaluate either the exact eigenangle-to-density Jacobian or the Gaussian holonomy fluctuation determinant, the mode integral is normalized relative to the pure-Haar integral. This removes the matter-independent saddle value and is separate from the removal of the angle-independent constant in the Fourier representation of the Haar measure. Including the vacuum contribution then gives
\begin{align}
\left.\log Z_{\rm crit,sing}\right|_{\rm saddle}
\simeq{}&
\frac12\sum_{n=1}^{\infty}\frac{A_n^2}{n}
+\frac12\sum_{n=1}^{\infty}\frac{A_{2n}}{n}
+\frac{N\beta r^2}{6}\int_0^\pi d\theta\,\sin\theta\,\sigma(\theta)^{3/2}.
\label{crit-logZ-onshell-raw}
\end{align}
Evaluating equation~\eqref{crit-logZ-onshell-raw} with equation~\eqref{crit-An-onshell} and the saddle profile in equation~\eqref{crit-sigma-profile} yields
\begin{align}
\left.\log Z_{\rm crit,sing}\right|_{\rm saddle}
\simeq{}&
\frac{2(Tr)^4}{(1-\omega^2)^2}
\Big[
\operatorname{Li}_5(e^{-2x})
+2x\operatorname{Li}_4(e^{-2x})
+x^2\operatorname{Li}_3(e^{-2x})
\Big]
\nonumber\\
&+\frac{(Tr)^2}{4(1-\omega^2)}
\Big[
\operatorname{Li}_3(e^{-2x})
+2x\operatorname{Li}_2(e^{-2x})
\Big]
+\frac{N(Tr)^2}{3(1-\omega^2)}x^3.
\label{crit-logZ-onshell}
\end{align}
The three terms in equation~\eqref{crit-logZ-onshell} arise, respectively, from the quadratic $A_n^2$ contribution, the one-body orthogonal-Haar term, and the Hubbard--Stratonovich vacuum term. Their relative sizes must be determined only after the gap equation has fixed $x$.

On the Gross--Witten--Wadia transition curve, $(Tr)^2/[N(1-\omega^2)]=O(N^0)$, and equation~\eqref{crit-gap-reduced} gives $x=O(N^0)$. Hence $A_n=O(N)$ for each fixed $n$. The first and third terms in equation~\eqref{crit-logZ-onshell} are $O(N^2)$, whereas the one-body Haar term is $O(N)$. At high temperature, the Gaussian functional trace of the Hubbard--Stratonovich field scales as $(Tr)^2/(1-\omega^2)=O(N)$ on this curve. It is consequently subleading to the $O(N^2)$ saddle and cannot modify the leading transition equations. This scaling is derived in equation~\eqref{app:total-fluctuation-scaling} of Appendix~\ref{app:sigma-determinant}.

A different ordering applies when $N\to\infty$ with $(Tr)^2/(1-\omega^2)$ fixed. In this limit the discussion following equation~\eqref{crit-gap-reduced} gives $x\to0$ and $A_n=O(N^0)$ for every fixed mode. Equation~\eqref{crit-holonomy-exact} then shows that the matter contribution and the even-mode Haar shift both enter $\rho_n$ at order $N^{-1}$. Accordingly, the first two terms in equation~\eqref{crit-logZ-onshell} and the Gaussian Hubbard--Stratonovich determinant all contribute at $O(N^0)$. Equations~\eqref{crit-logZ-onshell-raw} and~\eqref{crit-logZ-onshell} therefore give the on-shell saddle contribution, but not the complete $O(N^0)$ partition function.

The tangent-plane approximation additionally requires $Tr\gg1$, as quantified by equation~\eqref{app:relative-correction} in Appendix~\ref{app:tangent-validity}. Within this regime, take $N\to\infty$ first at fixed $Tr$ and fixed $|\omega|<1$, and only afterward take $|\omega|\to1$. Setting $x=0$ in equation~\eqref{crit-An-onshell} and substituting the result into the quadratic term of equation~\eqref{crit-logZ-onshell-raw} gives
\begin{equation}
\begin{aligned}
\left.\log Z_{\rm crit,sing}^{N=\infty}\right|_{\rm saddle}
&=\left.\frac12\sum_{n=1}^{\infty}\frac{A_n^2}{n}\right|_{x=0}
+O\!\left((1-\omega^2)^{-1}\right)
\\
&=\frac{2(Tr)^4}{(1-\omega^2)^2}
\sum_{n=1}^{\infty}\frac{1}{n^5}
+O\!\left((1-\omega^2)^{-1}\right)
\\
&=\frac{2\zeta(5)(Tr)^4}{(1-\omega^2)^2}
+O\!\left((1-\omega^2)^{-1}\right).
\end{aligned}
\label{crit-double-pole}
\end{equation}
Equation~\eqref{crit-double-pole} shows directly that the double pole comes from the $A_n^2$ term. Equation~\eqref{app:total-fluctuation-scaling} in Appendix~\ref{app:sigma-determinant} shows that the coupled one-loop correction grows only as $(Tr)^2/(1-\omega^2)$. It can therefore change the coefficient of the simple pole but cannot modify the double-pole term in equation~\eqref{crit-double-pole}.

The double-pole contribution also remains nonzero on the transition curve. Using equation~\eqref{crit-numerics}, for which $(Tr)^2/(1-\omega_c^2)=0.337640\,N$ and $x=x_c=0.644715$, the first term of equation~\eqref{crit-logZ-onshell} becomes
\begin{equation}
\begin{aligned}
\left.\frac12\sum_{n=1}^{\infty}\frac{A_n^2}{n}\right|_{c}
={}&2(0.337640)^2N^2
\Big[
\operatorname{Li}_5(e^{-2x_c})
+2x_c\operatorname{Li}_4(e^{-2x_c})
+x_c^2\operatorname{Li}_3(e^{-2x_c})
\Big]
\\
\simeq{}&0.172895\,N^2.
\end{aligned}
\label{crit-double-pole-transition}
\end{equation}
Combining equation~\eqref{crit-double-pole-transition} with the vacuum term in equation~\eqref{crit-logZ-onshell}, the transition-scale partition function has the structure
\begin{equation}
\begin{aligned}
\left.\log Z_{\rm crit,sing}\right|_c
&=\left.\frac12\sum_{n=1}^{\infty}\frac{A_n^2}{n}\right|_c
+\frac{0.337640}{3}x_c^3N^2+O(N)
\\
&\simeq0.203055\,N^2+O(N).
\end{aligned}
\label{crit-logZ-transition-scaling}
\end{equation}
The $O(N)$ terms in equation~\eqref{crit-logZ-transition-scaling} include the one-body Haar contribution and the Hubbard--Stratonovich one-loop determinant. Thus both the $A_n^2$ and vacuum terms contribute at leading order, whereas the one-loop determinant cannot modify the double-pole coefficient. At finite large $N$, the Gross--Witten--Wadia transition occurs before $|\omega|=1$ and cuts off the divergence. Equation~\eqref{crit-double-pole} therefore applies only to the ordered strict-$N$ limit in the ungapped phase.

\section{Discussion}
\label{sec:discussion}
We have studied the $O(N)$ model with the singlet constraint on $S^1\times S^2$ at both the free and critical fixed points in presence of an angular potential $\Omega$. At zero angular potential, the model undergoes a Gross-Witten-Wadia transition at $Tr\sim\sqrt{N}$ when the holonomy eigenvalue density develops a gap. The angular potential provides an additional parameter with which to tune this transition, as shown at fixed $Tr$ in equations~\eqref{intro-free-fixed-temperature} and~\eqref{intro-critical-fixed-temperature}. The curve $(Tr)^2\sim N(1-\Omega^2r^2)$ separates the ungapped and gapped phases in the $T$--$\omega$ plane.
The proportionality constant differs between the free and critical fixed points. The free singlet model admits the fixed-temperature large-$N$ result in equation~\eqref{intro-free-fixed-temperature} without a high-temperature approximation. For the critical singlet model, the auxiliary-field saddle is obtained for $Tr\gg1$, where $Tr$ is free to scale independent of $N$.

The interacting model in presence of a non-zero angular potential experiences breaking of $SO(3)$ symmetry to $SO(2)$. This leads to a polar angle dependent saddle for the auxiliary field path integral at the leading order in large $N$, unlike the case at zero angular potential, where only a constant mode of the auxiliary field is allowed at this order. Thus at non-zero angular potential the model is more involved. We solve for the saddle point of the auxiliary field $\sigma(\theta)$ in the leading tangent plane approximation at any given angle $\theta=\theta_0$. In this approximation we consider that at any point on $S^2$ with $\theta=\theta_0$, the auxiliary field remains constant on the tangent plane associated with that point. We find the solution for $\sigma(\theta_0)$ using this approximation and finally by combining the tangent patches at all the different values of $\theta$, we find the free energy at the leading order at high temperature. The corrections to the solution for $\sigma(\theta)$ apart from the tangent plane approximation lead to subleading effects on the free energy at high temperature.

The free energy in the ungapped phase exhibits a double pole as the leading contribution at $\Omega^2 r^2=1$ after the large-$N$ limit is taken. The model with higher spin symmetry \cite{Maldacena:2011jn,Maldacena:2012sf,Alba:2013yda,Alba:2015upa} in the singlet sector is non-local \cite{Aharony:2020omh}. Thus the leading simple pole in $(1-\Omega^2r^2)$ as predicted from the thermal effective action \cite{Anand:2025mfh,Advant:2026cqt,Cassani:2021fyv,Benjamin:2023qsc,Mukherjee:2026dfu} is not observed in this model. A model with a non-local interaction has also been studied in \cite{David:2026yis}, where the leading simple pole persists, but the non-analytic correction terms cause the free energy to deviate from the local effective field theory description. The study of our model on a pp wave geometry can provide an independent consistency check of the double pole obtained here, following the prescriptions given in \cite{Komargodski:2026ain}.

We observed that at a fixed temperature the Gross-Witten-Wadia transition can occur when $1-\Omega^2 r^2\sim \frac1N$. We again emphasize that the double pole nature is a property of the free energy in the ungapped phase. The $\Omega^2r^2=1$ behavior of the free energy right after the transition \cite{Amado:2016pgy} remains an interesting open question. It is expected that deep inside the gapped phase the model should resemble the unconstrained $O(N)$ model, thus should recover the simple pole \cite{Mishra:2026bpi} predicted by the thermal effective action.

Natural extensions include fermionic singlet models and theories with more
general interactions~\cite{Romatschke:2019ybu,Petkou:2018ynm,Grable:2022swa},
Chern--Simons matter theories~\cite{Giombi:2011kc,Aharony:2011jz}, singlet sector of the  models on squashed
spheres~\cite{Parmentier:2026aqh}, and $\mathcal{PT}$-symmetric models
\cite{Lawrence:2023woz,Grable:2023paf}. A more ambitious program is to study of the bulk dual of our model in the higher spin gravity in $AdS_4$.
The study of finite temperature properties of generic CFTs, such as thermal conformal blocks \cite{Buric:2024kxo,Buric:2025uqt,Ammon:2025cdz}, in presence of a non-zero angular potential can be an interesting direction in a related context. The study of CFT thermal one point functions in the fast rotation limit \cite{Miscioscia:2026bvl} remains an interesting direction to follow.

\section*{Acknowledgments}
The author thanks Justin David for useful discussions during the work and important comments on the final version of the manuscript.

ChatGPT and Codex have been used to perform algebraic manipulations, various consistency checks, aspects of the analysis and organization of the manuscript. Every step is human supervised and independently verified. 

The author is supported by the National Natural Science Foundation of China (NSFC) under Grants No. 12247103.
\appendix

\section{Orthogonal Haar measure at large $N$}
\label{app:haar-measure}

For $N=2K$, the Weyl Jacobian of the connected component is given by equation~\eqref{free-haar-jacobian}. Its logarithm contains the Fourier series of $\log|\sin(x/2)|$, which is only conditionally convergent. Multiplying each Fourier mode by $q^n$, with $0<q<1$, makes the series absolutely convergent. We thus define
\begin{equation}
L_q(x)=-\log 2-\sum_{n=1}^{\infty}\frac{q^n}{n}\cos(nx),
\qquad 0<q<1,
\qquad
\lim_{q\to1^-}L_q(x)=\log\left|\sin\frac{x}{2}\right|,
\label{app:haar-convergence-kernel}
\end{equation}
where the last equality holds for $x\notin2\pi\mathbb Z$. At fixed $q$, replacing each logarithm in equation~\eqref{free-haar-jacobian} by equation~\eqref{app:haar-convergence-kernel} gives
\begin{align}
\log J_{2K}^{(q)}
={}&-2K(K-1)\log2
-2\sum_{n=1}^{\infty}\frac{q^n}{n}
\sum_{a<b}\left[
\cos n(\alpha_a-\alpha_b)+\cos n(\alpha_a+\alpha_b)
\right] \notag\\
={}&-2K(K-1)\log2
-4\sum_{n=1}^{\infty}\frac{q^n}{n}
\sum_{a<b}\cos(n\alpha_a)\cos(n\alpha_b).
\label{app:haar-regulated-pair-sum}
\end{align}
The pair sum can be expressed in terms of traces by using
\begin{equation}
\sum_{a=1}^{K}\cos(n\alpha_a)=\frac12\operatorname{Tr}U^n,
\qquad
2\sum_{a<b}\cos(n\alpha_a)\cos(n\alpha_b)
=\left(\sum_{a=1}^{K}\cos(n\alpha_a)\right)^2
-\sum_{a=1}^{K}\cos^2(n\alpha_a),
\label{app:haar-pair-identities}
\end{equation}
together with
\begin{equation}
2\sum_{a=1}^{K}\cos^2(n\alpha_a)
=K+\frac12\operatorname{Tr}U^{2n}.
\label{app:haar-square-identity}
\end{equation}
Substitution into equation~\eqref{app:haar-regulated-pair-sum} yields
\begin{equation}
\log J_{2K}^{(q)}
=-\frac12\sum_{n=1}^{\infty}\frac{q^n}{n}(\operatorname{Tr}U^n)^2
+\frac12\sum_{n=1}^{\infty}\frac{q^n}{n}\operatorname{Tr}U^{2n}
+C_K(q),
\label{app:haar-regulated-traces}
\end{equation}
where
\begin{equation}
C_K(q)=-2K(K-1)\log2-K\log(1-q).
\label{app:haar-regulated-constant}
\end{equation}
Here we used $\sum_{n=1}^{\infty}q^n/n=-\log(1-q)$.

The apparent divergence of $C_K(q)$ is cancelled within the complete expression in equation~\eqref{app:haar-regulated-traces}. To display this cancellation exactly, first isolate the terms in $(\operatorname{Tr}U^n)^2$ for which the two eigenangle labels are equal:
\begin{equation}
4\sum_{a=1}^{K}\cos^2(n\alpha_a)
=2K+\operatorname{Tr}U^{2n}.
\label{app:haar-diagonal-trace}
\end{equation}
Their contribution to the first sum in equation~\eqref{app:haar-regulated-traces}, combined with its second sum, is
\begin{align}
&-\frac12\sum_{n=1}^{\infty}\frac{q^n}{n}
\left(2K+\operatorname{Tr}U^{2n}\right)
+\frac12\sum_{n=1}^{\infty}\frac{q^n}{n}\operatorname{Tr}U^{2n}
\notag\\
&\hspace{35mm}=-K\sum_{n=1}^{\infty}\frac{q^n}{n}
=K\log(1-q).
\label{app:haar-divergence-cancellation}
\end{align}
This cancels the term $-K\log(1-q)$ in equation~\eqref{app:haar-regulated-constant}. The limit $q\to1^-$ must therefore be taken only after the terms in equation~\eqref{app:haar-regulated-traces} have been combined. The remaining constant $-2K(K-1)\log2$ is independent of the eigenangles and is absorbed into the normalization $\mathcal N_K$. After this normalization, the finite holonomy-dependent part is
\begin{equation}
\log J_{2K}
=-\frac12\sum_{n=1}^{\infty}\frac{(\operatorname{Tr}U^n)^2}{n}
+\frac12\sum_{n=1}^{\infty}\frac{\operatorname{Tr}U^{2n}}{n},
\label{app:haar-finite-traces}
\end{equation}
with both sums defined using the common convergence factor $q^n$ and the limit $q\to1^-$, with the cancellation and normalization carried out as described above. This is equation~\eqref{free-haar-traces}. At coincident eigenangles the original Jacobian vanishes, so the divergence of its logarithm is physical and is not removed by this prescription.

This limiting procedure is distinct from the normalization of the large-$N$ mode integral. In the ungapped phase, combining equation~\eqref{app:haar-finite-traces} with the matter determinant and completing the square gives
\begin{align}
\log(J_{2K}Z_{\rm matter})
={}&-\frac12\sum_{n=1}^{\infty}\frac1n
\left[\operatorname{Tr}U^n-a_n-\frac{1+(-1)^n}{2}\right]^2
\notag\\
&+\frac12\sum_{n=1}^{\infty}\frac1n
\left[a_n+\frac{1+(-1)^n}{2}\right]^2.
\label{app:haar-square-completion}
\end{align}
The normalized Haar integral equals one when the matter source vanishes, namely when $a_n=0$. Because the present mode treatment does not include the exact angle-to-density Jacobian or the Gaussian holonomy determinant, its normalization is fixed by subtracting the saddle value at $a_n=0$. This gives
\begin{align}
\log Z_{O(N)}
&=\frac12\sum_{n=1}^{\infty}\frac1n
\left\{
\left[a_n+\frac{1+(-1)^n}{2}\right]^2
-\left[\frac{1+(-1)^n}{2}\right]^2
\right\}
\notag\\
&=\frac12\sum_{n=1}^{\infty}\frac{a_n^2+a_{2n}}{n}.
\label{app:haar-normalized-partition}
\end{align}
The first step, which regulates the Fourier series, and the second step, which normalizes the mode integral, are logically distinct. For $N=2K+1$, the Weyl measure again contains $K$ eigenangles, together with an additional one-body factor. The logarithm of this factor is $O(N)$ and therefore does not modify the leading $O(N^2)$ density saddle.

\section{Free singlet partition function from letter counting}
\label{app:letter-counting}

This appendix gives an independent derivation of the strict-$N=\infty$ result in equation~\eqref{free-strict-partition}. Here a letter is a state in the one-particle Hilbert space of a free conformal scalar. The derivation consists of counting these one-particle states, projecting a pair of them onto an $O(N)$ singlet, and then forming arbitrary products of the resulting singlet bilinears.

Introduce the thermal and angular fugacities
\begin{equation}
q=e^{-\beta/r},
\qquad
\eta=e^{\beta\Omega}.
\label{letter-fugacities}
\end{equation}
The condition $|\Omega r|<1$ ensures that $q\eta<1$ and $q/\eta<1$, so all sums below converge. Denote the scalar one-particle Hilbert space by $\mathcal H_S$. Its states are labelled by $(\ell,m)$, with
\begin{equation}
E_\ell=\frac{\ell+\frac12}{r},
\qquad
m=-\ell,-\ell+1,\ldots,\ell,
\qquad
\ell=0,1,2,\ldots .
\label{letter-scalar-spectrum}
\end{equation}
The weight of such a state is $q^{rE_\ell}\eta^m=q^{\ell+1/2}\eta^m$. The one-particle character is consequently
\begin{equation}
z_S(q,\eta)
\equiv\operatorname{Tr}_{\mathcal H_S}\!\left(q^{rH}\eta^{J_z}\right)
=\sum_{\ell=0}^{\infty}q^{\ell+1/2}
\sum_{m=-\ell}^{\ell}\eta^m
=\frac{q^{1/2}(1+q)}{(1-q\eta)(1-q/\eta)}.
\label{letter-scalar-character}
\end{equation}
Equivalently, before cancelling a factor of $1-q$, this character can be written as $q^{1/2}(1-q^2)/[(1-q)(1-q\eta)(1-q/\eta)]$. In this form the three denominator factors count translation descendants with angular momenta $0,+1,-1$, while $1-q^2$ removes the level-two null state imposed by the free equation of motion.

We next impose the singlet condition. Let $A$ and $B$ denote states in $\mathcal H_S$. Contracting the two vector indices with the invariant tensor $\delta_{ij}$ gives the bilinear
\begin{equation}
\mathcal O_{AB}=\sum_{i=1}^{N}\phi_{i,A}\phi_{i,B},
\qquad
\mathcal O_{AB}=\mathcal O_{BA}.
\label{letter-singlet-bilinear}
\end{equation}
Because the scalar letters are bosonic, the space of bilinear singlets is the symmetric square $\operatorname{Sym}^2\mathcal H_S$. Let $g=q^{rH}\eta^{J_z}$, choose a basis of simultaneous eigenstates $|A\rangle$ of $H$ and $J_z$, and denote the weight of $|A\rangle$ by $w_A=q^{rE_A}\eta^{m_A}$. The relation $\mathcal O_{AB}=\mathcal O_{BA}$ means that the pairs $(A,B)$ and $(B,A)$ represent the same bilinear. Thus a pair of distinct states is counted once, while the diagonal pair $(A,A)$ is also retained once. The bilinear character is therefore
\begin{equation}
\begin{aligned}
z_{\rm bil}(q,\eta)
&=\operatorname{Tr}_{\operatorname{Sym}^2\mathcal H_S}(g\otimes g)
=\sum_{A<B}w_Aw_B+\sum_Aw_A^2
\\
&=\frac12\left[\left(\sum_Aw_A\right)^2+\sum_Aw_A^2\right]
\\
&=\frac12\left[z_S(q,\eta)^2+z_S(q^2,\eta^2)\right].
\end{aligned}
\label{letter-bilinear-character}
\end{equation}
The second line follows by expanding
$\left(\sum_Aw_A\right)^2=\sum_Aw_A^2+2\sum_{A<B}w_Aw_B$.
For the final line, $\sum_Aw_A=z_S(q,\eta)$, while
$\sum_Aw_A^2=\sum_Aq^{2rE_A}\eta^{2m_A}=z_S(q^2,\eta^2)$; squaring a weight doubles both the energy and the angular momentum. Equivalently, equation~\eqref{letter-bilinear-character} follows by inserting the symmetric projector $(1+P_{12})/2$ in $\mathcal H_S\otimes\mathcal H_S$. The identity part gives $z_S(q,\eta)^2$, whereas the exchange part identifies the two state labels and gives $z_S(q^2,\eta^2)$. This exchange contribution is the origin of the $a_{2n}$ term in the final partition function.

At strict $N=\infty$, the finite-$N$ algebraic relations among products of bilinears are absent at every fixed operator degree. The full singlet spectrum is therefore the bosonic Fock space generated by the bilinear states. Its partition function is the plethystic exponential of equation~\eqref{letter-bilinear-character},
\begin{equation}
Z_{\rm sing}^{N=\infty}(q,\eta)
=\exp\left[
\sum_{n=1}^{\infty}\frac{1}{n}
z_{\rm bil}(q^n,\eta^n)
\right].
\label{letter-plethystic}
\end{equation}
The sum over $n$ allows each bosonic bilinear state to occur with arbitrary occupation number. The coefficient $a_n$ introduced in the matrix-integral calculation is
\begin{equation}
a_n
=\sum_{\ell=0}^{\infty}\sum_{m=-\ell}^{\ell}
q^{n(\ell+1/2)}\eta^{nm}
=z_S(q^n,\eta^n),
\label{letter-an-character}
\end{equation}
so that $z_S(q^{2n},\eta^{2n})=a_{2n}$. Substitution of equation~\eqref{letter-bilinear-character} into equation~\eqref{letter-plethystic} then gives
\begin{align}
\log Z_{\rm sing}^{N=\infty}
&=\frac12\sum_{n=1}^{\infty}\frac{1}{n}
\left[z_S(q^n,\eta^n)^2+z_S(q^{2n},\eta^{2n})\right]
\nonumber\\
&=\frac12\sum_{n=1}^{\infty}\frac{a_n^2+a_{2n}}{n}.
\label{letter-final-partition}
\end{align}
This is precisely equation~\eqref{free-strict-partition}. The holonomy-independent vacuum-energy factor has been omitted in both derivations. At finite $N$, sufficiently high products of bilinears obey algebraic relations, and the unrestricted plethystic exponential must be modified. These relations do not affect the strict large-$N$ result at any fixed operator degree.
\section{Local tangent-plane integrals}
\label{app:tangent-integrals}

This appendix evaluates the momentum integral in the first line of equation~\eqref{crit-winding-coefficient}. At fixed latitude $\theta$, use polar coordinates in the local momentum plane. We denote the momentum-space polar angle by $\psi$ to distinguish it from the latitude, so that
$p_y=p\sin\psi$ and $d^2p=p\,dp\,d\psi$. Suppressing the $\theta$ dependence of $v$, $\Gamma=(1-v^2)^{-1/2}$, and $\sigma$, the integral becomes
\begin{align}
	\int\frac{d^2p}{(2\pi)^2}
	e^{-n\beta[\sqrt{p^2+\sigma}-vp_y]}
	&=\frac{1}{(2\pi)^2}\int_0^\infty p\,dp\,
	e^{-n\beta\sqrt{p^2+\sigma}}
	\int_0^{2\pi}d\psi\,e^{n\beta vp\sin\psi}
	\nonumber\\
	&=\frac{1}{2\pi}\int_0^\infty p\,dp\,
	e^{-n\beta\sqrt{p^2+\sigma}}I_0(n\beta vp)
	\nonumber\\
	&=\frac{e^{-n\beta\sqrt{\sigma}/\Gamma}}{2\pi}
	\left[
	\frac{\sqrt{\sigma}\,\Gamma^2}{n\beta}
	+\frac{\Gamma^3}{(n\beta)^2}
	\right].
	\label{app:tangent-momentum-result}
\end{align}
Here $I_0$ is the modified Bessel function, and the second line follows from
$\int_0^{2\pi}d\psi\,e^{z\sin\psi}=2\pi I_0(z)$. To evaluate the radial integral, first introduce a Schwinger parameter $s$ through
\begin{equation}
	e^{-a\sqrt{p^2+M^2}}
	=\frac{a}{2\sqrt{\pi}}
	\int_0^\infty\frac{ds}{s^{3/2}}\,
	\exp\left[-s(p^2+M^2)-\frac{a^2}{4s}\right],
	\qquad a>0.
	\label{app:schwinger-square-root}
\end{equation}
The remaining momentum integral is the standard integral formula~(10.43.23) of reference~\cite{NIST:DLMF}, after setting $\nu=0$,
\begin{equation}
	\int_0^\infty p\,dp\,e^{-sp^2}I_0(bp)
	=\frac{1}{2s}\exp\left(\frac{b^2}{4s}\right),
	\qquad s>0.
	\label{app:gaussian-bessel-identity}
\end{equation}
Combining equations~\eqref{app:schwinger-square-root} and~\eqref{app:gaussian-bessel-identity}, and then performing the $s$ integral, gives
\begin{equation}
	\int_0^\infty p\,dp\,
	e^{-a\sqrt{p^2+M^2}}I_0(bp)
	=
	\frac{a\left(1+M\sqrt{a^2-b^2}\right)}
	{\left(a^2-b^2\right)^{3/2}}\,
	e^{-M\sqrt{a^2-b^2}},
	\qquad a>|b|,\quad M\geq0.
	\label{app:bessel-radial-identity}
\end{equation}
In the present case, $a=n\beta$, $b=n\beta v$, and $M=\sqrt{\sigma}$. Since $|v|<1$,
\begin{equation}
	\sqrt{a^2-b^2}
	=n\beta\sqrt{1-v^2}
	=\frac{n\beta}{\Gamma},
	\label{app:bessel-parameter-reduction}
\end{equation}
and equation~\eqref{app:bessel-radial-identity} gives the last line of equation~\eqref{app:tangent-momentum-result}. Restoring the $\theta$ dependence, using $\beta=1/T$, and integrating over the sphere gives
\begin{align}
	&2\pi r^2\int_0^\pi d\theta\,\sin\theta
	\int\frac{d^2p}{(2\pi)^2}
	e^{-n\beta[\sqrt{p^2+\sigma(\theta)}-v(\theta)p_y]}
	\nonumber\\
	&=(Tr)^2\int_0^\pi d\theta\,\sin\theta\,
	e^{-n\sqrt{\sigma(\theta)}/[T\Gamma(\theta)]}
	\left[
	\frac{\Gamma(\theta)^3}{n^2}
	+\frac{\sqrt{\sigma(\theta)}}{T}\frac{\Gamma(\theta)^2}{n}
	\right],
	\label{app:tangent-winding-result}
\end{align}
which is precisely the second line of equation~\eqref{crit-winding-coefficient}.

\section{High-$T$ expansion and the tangent-plane approximation}
\label{app:tangent-validity}

This appendix connects the local calculation in
section~\ref{sec:critical-model} to the exact determinant on $S^2$.
Starting from equation~\eqref{crit-effective-action}, we show that the
coefficient $A_n[\sigma]$ defined in
equation~\eqref{crit-winding-coefficient} is the leading
high-temperature term of the exact spherical coefficient. This
derivation does not assume the saddle-point solution for
$\sigma(\theta)$.

For the axisymmetric background in
equation~\eqref{crit-axisymmetric-sigma}, define
\begin{equation}
	H_\sigma
	=
	\sqrt{
		-\nabla_{S^2_r}^2+\frac{R}{8}+\sigma(\theta)
	}.
	\label{app:tangent-spatial-operator}
\end{equation}
Since $[H_\sigma,L_z]=0$, performing the Matsubara sum and expanding
$-\log(1-z)=\sum_{n\geq1}z^n/n$ gives
\begin{equation}
	\log Z_{a,\mathrm{th}}
	=
	2\sum_{n=1}^{\infty}
	\frac{\cos(n\alpha_a)}{n}\,
	{\cal A}_n[\sigma],
	\label{app:exact-thermal-winding-expansion}
\end{equation}
where the exact spherical coefficient is
\begin{equation}
	{\cal A}_n[\sigma]
	=
	{\rm Tr}_{S^2_r}
	\exp\left[
	-n\beta\left(
	H_\sigma-\Omega L_z
	\right)
	\right].
	\label{app:exact-winding}
\end{equation}

Introduce
\begin{equation}
	h=\frac{\beta}{r}=\frac{1}{Tr},
	\qquad
	\omega=\Omega r,
	\qquad
	\widehat{\sigma}(\theta)=\beta^2\sigma(\theta).
	\label{app:tangent-scaled-variables}
\end{equation}
Using $R=2/r^2$ and the variables in
equation~\eqref{app:tangent-scaled-variables},
equation~\eqref{app:exact-winding} becomes
\begin{equation}
	{\cal A}_n[\sigma]
	=
	{\rm Tr}_{S^2}e^{-nP_h},
	\qquad
	P_h
	=
	\sqrt{
		-h^2\Delta_{S^2}
		+\widehat{\sigma}(\theta)
		+\frac{h^2}{4}
	}
	-h\omega L_z .
	\label{app:semiclassical-operator}
\end{equation}
Thus the high-temperature limit $Tr\gg1$ is the semiclassical limit
$h\to0$.

Around each point on the sphere, use the local orthonormal coordinates
$(x,y)$ of equation~\eqref{crit-local-coordinates}, with conjugate
dimensionless momenta $(q_x,q_y)$. The principal symbol of
equation~\eqref{app:semiclassical-operator} is
\begin{equation}
	p_0
	=
	\sqrt{
		q_x^2+q_y^2+\widehat{\sigma}(\theta)
	}
	-\omega\sin\theta\,q_y .
	\label{app:principal-symbol}
\end{equation}
The term $h^2/4$ is explicitly of order $h^2$. In the invariant
half-density formulation, $-h^2\Delta_{S^2}$ has vanishing
subprincipal symbol, while multiplication by
$\widehat{\sigma}(\theta)$ introduces no subprincipal term.
Semiclassical functional calculus therefore implies that the
square-root operator in equation~\eqref{app:semiclassical-operator}
also has vanishing subprincipal symbol. Finally, $L_z$ generates a
divergence-free Killing vector and likewise has vanishing subprincipal
symbol. Consequently $P_h$ has no term linear in $h$.

Standard semiclassical functional calculus and the Weyl trace formula
\cite{DimassiSjostrand1999} then give, for fixed $n$ and fixed
$|\omega|<1$,
\begin{align}
	{\cal A}_n[\sigma]
	={}&
	\frac{1}{h^2}
	\int_{S^2}d\Omega
	\int\frac{d^2q}{(2\pi)^2}
	\exp\left[
	-n\left(
	\sqrt{q^2+\widehat{\sigma}(\theta)}
	-\omega\sin\theta\,q_y
	\right)
	\right]
	+O(1)
	\nonumber\\
	={}&
	(Tr)^2
	\int_{S^2}d\Omega
	\int\frac{d^2q}{(2\pi)^2}
	\exp\left[
	-n\left(
	\sqrt{q^2+\frac{\sigma(\theta)}{T^2}}
	-\omega\sin\theta\,q_y
	\right)
	\right]
	+O(1).
	\label{app:weyl-highT}
\end{align}
The remaining integral is dimensionless. The leading contribution is
therefore of order $(Tr)^2$, whereas the first correction is of order
$(Tr)^0$. Hence the relative correction is
\begin{equation}
	O\left(\frac{1}{(Tr)^2}\right)
	=
	O\left(\frac{\beta^2}{r^2}\right).
	\label{app:relative-correction}
\end{equation}

Finally, returning to physical momenta through $q_i=\beta p_i$ and
using $dA=r^2d\Omega$, equation~\eqref{app:weyl-highT} becomes
\begin{equation}
	{\cal A}_n[\sigma]
	=
	\int_{S^2_r}dA
	\int\frac{dp_x\,dp_y}{(2\pi)^2}
	\exp\left[
	-n\beta\left(
	\sqrt{p_x^2+p_y^2+\sigma(\theta)}
	-\Omega r\sin\theta\,p_y
	\right)
	\right]
	+O(1).
	\label{app:tangent-from-weyl}
\end{equation}
The integral in equation~\eqref{app:tangent-from-weyl} is precisely the
first line of equation~\eqref{crit-winding-coefficient}. Consequently,
\begin{equation}
	{\cal A}_n[\sigma]
	=
	A_n[\sigma]+O(1),
	\qquad
	Tr\longrightarrow\infty,
	\label{app:exact-local-matching}
\end{equation}
for fixed $n$ and fixed $|\omega|<1$.
Equations~\eqref{app:relative-correction} and
\eqref{app:exact-local-matching} establish that the tangent-plane
calculation gives the leading high-temperature contribution to the
exact spherical determinant.

For $\sigma=0$, equation~\eqref{app:exact-local-matching} gives
\begin{equation}
	{\cal A}_n[0]
	=
	\frac{2(Tr)^2}{n^2(1-\omega^2)}
	+O(1),
	\label{app:tangent-free-check}
\end{equation}
in agreement with the direct high-temperature expansion in
equation~\eqref{free-an-high-temperature}.

The corresponding local Weyl expansion also applies to the
dimensionally regularized holonomy-independent vacuum contribution in
equation~\eqref{crit-vacuum-contribution}. For the $K=N/2$ complex
scalars, the exact spherical vacuum term is
\begin{equation}
	\log Z_0[\sigma]
	=
	-\frac{N\beta}{2}\,
	{\rm Tr}_{S^2_r} H_\sigma .
	\label{app:exact-vacuum-trace}
\end{equation}
Using the variables in equation~\eqref{app:tangent-scaled-variables},
this becomes
\begin{equation}
	\log Z_0[\sigma]
	=
	-\frac{N}{2}\,
	{\rm Tr}_{S^2}
	\sqrt{
		-h^2\Delta_{S^2}
		+\widehat{\sigma}(\theta)
		+\frac{h^2}{4}
	}.
	\label{app:scaled-vacuum-trace}
\end{equation}
Applying the corresponding local Weyl expansion to this regularized
vacuum trace gives
\begin{align}
	\log Z_0[\sigma]
	=
	N\left[
	-\frac{1}{2h^2}
	\int_{S^2}d\Omega
	\int\frac{d^2q}{(2\pi)^2}
	\sqrt{q^2+\widehat{\sigma}(\theta)}
	+O(h^0)
	\right],
	\label{app:vacuum-weyl}
\end{align}
where the momentum integral is understood in the same dimensional
regularization as in the main text. The required integral is
\begin{equation}
	\int\frac{d^2q}{(2\pi)^2}
	\sqrt{q^2+\widehat{\sigma}}
	=
	-\frac{\widehat{\sigma}^{3/2}}{6\pi}.
	\label{app:vacuum-momentum-integral}
\end{equation}
Combining equation~\eqref{app:vacuum-momentum-integral} with
$\widehat{\sigma}=\beta^2\sigma$ and $h=\beta/r$, the leading term is
\begin{equation}
	\log Z_0[\sigma]
	=
	\frac{N\beta}{12\pi}
	\int_{S^2_r}dA\,\sigma(\theta)^{3/2}
	+
	N\,O(h^0).
	\label{app:vacuum-local-leading}
\end{equation}
Thus the leading term of equation~\eqref{app:vacuum-local-leading} is
equation~\eqref{crit-vacuum-contribution}. For
$\widehat{\sigma}=O(1)$, the correction is suppressed relative to the
leading term by
\begin{equation}
	O(h^2)
	=
	O\left(\frac{\beta^2}{r^2}\right).
	\label{app:vacuum-relative-correction}
\end{equation}

It remains to check a posteriori that the resulting saddle varies
slowly over the local thermal scale. Using
$\sigma(\theta)=T^2x^2\Gamma(\theta)^2$, which follows from the saddle
profile in equation~\eqref{crit-sigma-profile}, one finds, for $\theta$
close to $\theta_0$,
\begin{equation}
	\frac{\beta^2[\sigma(\theta)-\sigma(\theta_0)]}
	{\Gamma(\theta_0)^2}
	=
	x^2\left[
	\frac{\Gamma(\theta)^2}{\Gamma(\theta_0)^2}-1
	\right]
	=
	2x^2\frac{\Gamma(\theta)-\Gamma(\theta_0)}
	{\Gamma(\theta_0)}
	+O\!\left((\theta-\theta_0)^2\right).
	\label{app:tangent-saddle-expansion}
\end{equation}
Since the local thermal angular scale is
\begin{equation}
	|\theta-\theta_0|
	=
	O\!\left(
	\frac{\beta}{r\Gamma(\theta_0)}
	\right),
	\label{app:local-thermal-angular-scale}
\end{equation}
the dimensionless variation of the leading saddle is
\begin{equation}
	\frac{\beta^2|\sigma(\theta)-\sigma(\theta_0)|}
	{\Gamma(\theta_0)^2}
	=
	O\!\left(x^2\frac{\beta}{r}\right)
	=
	O\!\left(\frac{\beta}{r}\right)
	\ll1 ,
	\label{app:tangent-saddle-consistency}
\end{equation}
for $Tr\gg1$, fixed $|\omega|<1$, and $x=O(1)$ or smaller.  At $x=0$ the
left-hand side vanishes identically. Thus the saddle obtained from the
leading local determinant is self-consistent, including the strict ungapped
limit.

\paragraph{Uniformity in the fast-rotation limit.}

The Weyl expansion above was derived for fixed $|\omega|<1$. We now
examine its behavior in the correlated limit $\omega\to1$ relevant for
the rotation-driven transition. We take $\omega\to1^{-}$ for
definiteness; the limit $\omega\to-1$ follows by reversing the sign of
the azimuthal angular momentum.

Let
\begin{equation}
	\varepsilon=1-\omega^2 .
	\label{app:fast-epsilon}
\end{equation}
The leading phase-space result already shows that the fast-rotation
singularity is localized near the equator. Indeed, writing
\begin{equation}
	\theta=\frac{\pi}{2}+s ,
\end{equation}
one finds, for $\varepsilon\ll1$ and $|s|\ll1$,
\begin{equation}
	1-\omega^2\sin^2\theta
	=
	\varepsilon+s^2
	+O(\varepsilon s^2,s^4).
\end{equation}
Consequently,
\begin{equation}
	\Gamma(\theta)^3
	\simeq
	\frac{1}{(\varepsilon+s^2)^{3/2}} .
\end{equation}
The singular contribution therefore arises from an equatorial region
of width
\begin{equation}
	|s|\sim\sqrt{\varepsilon}.
\end{equation}
It is natural to resolve this region by introducing
\begin{equation}
	\theta
	=
	\frac{\pi}{2}
	+\sqrt{\varepsilon}\,u .
	\label{app:fast-equatorial-coordinate}
\end{equation}
Then
\begin{equation}
	1-\omega^2\sin^2\theta
	=
	\varepsilon(1+u^2)
	+O(\varepsilon^2),
	\qquad
	\Gamma^2(\theta)
	=
	\frac{1}{\varepsilon(1+u^2)}
	\left[1+O(\varepsilon)\right].
	\label{app:fast-gamma}
\end{equation}
For example,
\begin{equation}
	d\theta\,\Gamma^3
	\sim
	\frac{1}{\varepsilon}
	\frac{du}{(1+u^2)^{3/2}},
\end{equation}
which directly reproduces the $1/(1-\omega^2)$ enhancement of the
leading coefficient.

We now examine the exact semiclassical operator in
equation~\eqref{app:semiclassical-operator} in this region. On the
leading saddle,
\begin{equation}
	\widehat{\sigma}(\theta)
	=
	\beta^2\sigma(\theta)
	=
	x^2\Gamma(\theta)^2 .
	\label{app:fast-sigmahat}
\end{equation}
Decompose into azimuthal modes,
\begin{equation}
	L_z=m.
\end{equation}
As $\omega\to1$, the singular sector consists of large positive,
co-rotating angular momenta. Near the equator, a large co-rotating mode
has a dimensionless residual rotating energy of order
\begin{equation}
	\beta(E-\Omega m)
	\sim
	hm(1-\omega)
	\sim
	\frac12 hm\varepsilon .
\end{equation}
The modes that remain thermally relevant therefore satisfy
\begin{equation}
	hm\varepsilon=O(1),
	\qquad\text{or equivalently}\qquad
	m\sim\frac{1}{h\varepsilon}.
	\label{app:fast-m-scaling}
\end{equation}
We accordingly introduce the finite variable
\begin{equation}
	\kappa=h\varepsilon m .
	\label{app:fast-kappa}
\end{equation}

Using equations~\eqref{app:fast-equatorial-coordinate} and
\eqref{app:fast-kappa}, the terms entering $P_h$ scale as
\begin{align}
	\frac{h^2m^2}{\sin^2\theta}
	&=
	\frac{\kappa^2}{\varepsilon^2}
	+\frac{\kappa^2u^2}{\varepsilon}
	+O(1),
	\nonumber\\
	-h^2\partial_\theta^2
	&=
	-\frac{h^2}{\varepsilon}\partial_u^2,
	\nonumber\\
	\widehat{\sigma}(\theta)
	&=
	\frac{x^2}{\varepsilon(1+u^2)}
	+O(1).
	\label{app:fast-operator-scaling}
\end{align}
Since
\begin{equation}
	\omega
	=
	1-\frac{\varepsilon}{2}
	+O(\varepsilon^2),
\end{equation}
the leading $O(\varepsilon^{-1})$ pieces of the square-root energy and
the rotational term cancel. The resulting boundary-layer operator is
\begin{equation}
	P_{h,m}
	=
	\frac{\kappa}{2}(1+u^2)
	+
	\frac{1}{2\kappa}
	\left[
	-h^2\partial_u^2
	+\frac{x^2}{1+u^2}
	\right]
	+O(\varepsilon).
	\label{app:fast-rotation-operator}
\end{equation}
The important point is that the semiclassical parameter of the
resulting one-dimensional problem remains $h=\beta/r$: no inverse
power of $\varepsilon$ multiplies $h$.

Adjacent azimuthal modes correspond to
\begin{equation}
	\Delta\kappa=h\varepsilon ,
\end{equation}
and therefore, in the singular co-rotating sector,
\begin{equation}
	\sum_m
	\longrightarrow
	\frac{1}{h\varepsilon}
	\int_0^\infty d\kappa .
	\label{app:fast-m-sum}
\end{equation}
For each fixed $\kappa$, the remaining $u$-trace has the usual
one-dimensional Weyl measure $(2\pi h)^{-1}du\,dp$. Consequently, for
fixed $n$,
\begin{equation}
	{\cal A}_n
	=
	\frac{1}{h^2\varepsilon}
	\left[
	C_{0,n}(x)
	+h^2C_{2,n}(x)
	+O(h^4)
	+O(\varepsilon)
	\right],
	\label{app:fast-rotation-weyl}
\end{equation}
where $C_{0,n}(x)$ is the phase-space contribution obtained from the
principal symbol of equation~\eqref{app:fast-rotation-operator}. If
${\cal A}_n^{(0)}$ denotes this leading boundary-layer contribution,
then
\begin{equation}
	\frac{
		{\cal A}_n-{\cal A}_n^{(0)}
	}{
		{\cal A}_n^{(0)}
	}
	=
	O(h^2)+O(\varepsilon).
	\label{app:fast-relative-correction}
\end{equation}
Here the $O(h^2)$ term is the genuine semiclassical correction, whereas
the $O(\varepsilon)$ term represents the error made in truncating the
fast-rotation boundary-layer expansion itself.

Thus the enhancement
\begin{equation}
	{\cal A}_n
	\sim
	\frac{1}{h^2(1-\omega^2)}
	=
	\frac{(Tr)^2}{1-\omega^2}
	\label{app:fast-winding-scaling}
\end{equation}
comes entirely from the enlarged equatorial phase space, while the
relative semiclassical corrections remain controlled by
$h^2=(Tr)^{-2}$. In particular, no additional condition relating
$1-\omega^2$ to $h$ is required for the leading boundary-layer
expansion. On the transition curve $x=x_c>0$, the coefficients are
exponentially suppressed at large $n$, so there is no additional
large-$n$ enhancement competing with this high-temperature counting.

The same order-of-magnitude counting applies to the vacuum
contribution. Although $\widehat{\sigma}=x^2\Gamma^2$ grows as
$1/\varepsilon$ in the equatorial region, its leading Weyl term scales
as
\begin{equation}
	\log Z_0^{(0)}
	\sim
	N h^{-2}
	\int d\theta\,
	\widehat{\sigma}^{3/2}
	=
	O\!\left(
	\frac{N}{h^2\varepsilon}
	\right).
	\label{app:fast-vacuum-leading}
\end{equation}
The first local corrections contain one curvature or two derivatives
of the slowly varying background. Under
$\theta-\pi/2=\sqrt{\varepsilon}\,u$, their most singular angular
integrals are at most $O(\varepsilon^{-1})$. Since these terms do not
carry the leading factor $h^{-2}$, their ratio to
equation~\eqref{app:fast-vacuum-leading} remains
\begin{equation}
	\frac{\delta\log Z_0}
	{\log Z_0^{(0)}}
	=
	O(h^2)
	=
	O\!\left(\frac{1}{(Tr)^2}\right).
	\label{app:fast-vacuum-relative}
\end{equation}
Thus both the thermal and vacuum contributions entering the leading
saddle remain uniformly controlled in the correlated high-temperature
and fast-rotation limit.

Finally, these corrections also feed back through the holonomy saddle.
In the following two equations, $O(h^2+\varepsilon)$ denotes a relative
remainder in the joint limit $h\to0$ and $\varepsilon\to0$. No hierarchy
between the two parameters is assumed: the $O(h^2)$ part is the
semiclassical correction, while the $O(\varepsilon)$ part is the
fast-rotation boundary-layer correction. Thus, for each fixed $n$,
\begin{equation}
	{\cal A}_n
	=
	{\cal A}_n^{(0)}
	\left[
	1+O(h^2+\varepsilon)
	\right],
	\label{app:fast-An-correction}
\end{equation}
equation~\eqref{crit-holonomy-exact} gives
\begin{equation}
	\rho_n
	=
	\frac{{\cal A}_n^{(0)}}{N}
	\left[
	1+O(h^2+\varepsilon)
	\right]
	+
	\frac{1+(-1)^n}{2N}.
	\label{app:fast-rho-correction}
\end{equation}
On the Gross--Witten--Wadia transition curve,
${\cal A}_n^{(0)}=O(N)$ for each fixed $n$, and hence the correction to the
leading matter contribution to $\rho_n$ is of relative order
$O(h^2+\varepsilon)$. Since $x=x_c>0$ on the transition curve, the
sums over $n$ entering the auxiliary-field gap equation are
exponentially convergent. The direct thermal and vacuum corrections,
together with their indirect feedback through the corrected holonomy
moments, therefore shift the solution only parametrically as
\begin{equation}
	x_c
	=
	x_c^{(0)}
	+
	O(h^2)
	+
	O(\varepsilon),
\end{equation}
and similarly the critical relation takes the form
\begin{equation}
	\frac{(Tr)^2}
	{N(1-\omega_c^2)}
	=
	C_c^{(0)}
	\left[
	1
	+O\!\left(\frac{1}{(Tr)^2}\right)
	+O(1-\omega_c^2)
	\right],
	\label{app:critical-relation-corrections}
\end{equation}
where $C_c^{(0)}$ denotes the value obtained from the leading local
calculation. Thus the curvature and derivative corrections, including
their feedback through the holonomy saddle, do not modify the leading
high-temperature Gross--Witten--Wadia transition equations.

\section{High-$T$ scaling of the one-loop determinant}
\label{app:sigma-determinant}

The saddle-point result in equation~\eqref{crit-logZ-onshell} does not
include fluctuations about the coupled $(\sigma,\rho_n)$ saddle.  In
this appendix we do not attempt to compute their complete finite
contribution.  Our purpose is only to determine their parametric order
at high temperature and to show that they cannot modify either the
leading large-$N$ transition equations or the double-pole term obtained
from the saddle.

It is useful first to make explicit the role of the holonomy modes.
Define
\begin{equation}
	b_n\equiv\frac{1+(-1)^n}{2}.
\end{equation}
The part of the exponent in
equation~\eqref{crit-large-N-functional} that depends on $\rho_n$ can
be completed into a square,
\begin{align}
	{\cal W}[\rho,\sigma]
	={}&
	-\frac{N^2}{2}
	\sum_{n=1}^{\infty}\frac{1}{n}
	\left[
	\rho_n-\frac{A_n[\sigma]+b_n}{N}
	\right]^2
	\nonumber\\
	&+
	\frac12
	\sum_{n=1}^{\infty}
	\frac{\left[A_n[\sigma]+b_n\right]^2}{n}
	+
	N V_0[\sigma],
	\label{app:coupled-square}
\end{align}
where
\begin{equation}
	V_0[\sigma]
	=
	\frac{\beta r^2}{6}
	\int_0^\pi d\theta\,
	\sin\theta\,\sigma(\theta)^{3/2}.
\end{equation}
The centre of the Gaussian in the holonomy directions is therefore
\begin{equation}
	\rho_n^*
	=
	\frac{A_n[\sigma]+b_n}{N},
	\label{app:rho-star}
\end{equation}
which is precisely the saddle in equation~\eqref{crit-holonomy-exact}.

The quadratic Hessian of the first line of
equation~\eqref{app:coupled-square} with respect to $\rho_n$ is
proportional to $N^2/n$ and is independent of $\sigma$, temperature and
angular potential.  Its Gaussian determinant is consequently part of
the matter-independent holonomy normalization discussed in
section~\ref{sec:critical-free-energy}.  The non-trivial dependence on
the Hubbard--Stratonovich field is contained in the reduced functional
obtained after imposing the holonomy saddle,
\begin{equation}
	{\cal W}_{\rm red}[\sigma]
	=
	\frac12
	\sum_{n=1}^{\infty}
	\frac{\left[A_n[\sigma]+b_n\right]^2}{n}
	+
	N V_0[\sigma].
	\label{app:reduced-sigma-functional}
\end{equation}

Since the variables $\rho_n$ are global holonomy modes, their quadratic
mixing with the Hubbard--Stratonovich field can involve only the
static, axisymmetric component of $\sigma$.  We display this sector
explicitly below.  Hubbard--Stratonovich fluctuations with nonzero
Matsubara frequency or nonzero azimuthal momentum do not mix with
$\rho_n$ and are included in the local fluctuation determinant
discussed subsequently.

For the static axisymmetric sector, the second variation of
equation~\eqref{app:reduced-sigma-functional} has the schematic form
\begin{align}
	K_{\rm eff}(\theta,\theta')
	={}&
	K_{\rm loc}(\theta,\theta')
	+
	K_\rho(\theta,\theta'),
	\label{app:kernel-decomposition}
	\\
	K_{\rm loc}(\theta,\theta')
	={}&
	\sum_{n=1}^{\infty}
	\frac{A_n+b_n}{n}
	\frac{\delta^2 A_n}
	{\delta\sigma(\theta)\delta\sigma(\theta')}
	+
	N
	\frac{\delta^2V_0}
	{\delta\sigma(\theta)\delta\sigma(\theta')},
	\label{app:local-kernel}
	\\
	K_\rho(\theta,\theta')
	={}&
	\sum_{n=1}^{\infty}
	\frac{1}{n}
	\frac{\delta A_n}{\delta\sigma(\theta)}
	\frac{\delta A_n}{\delta\sigma(\theta')}.
	\label{app:rho-induced-kernel}
\end{align}
The last term is the correction induced by the motion of the holonomy
saddle when $\sigma$ is varied.  Equivalently, it is the
Schur-complement contribution obtained after eliminating the Gaussian
holonomy fluctuations.  Thus the feedback of the $\rho_n$ saddle is
included in $K_{\rm eff}$.

We now determine the high-temperature order of the two contributions
in equation~\eqref{app:kernel-decomposition}.

\paragraph{Local Hubbard--Stratonovich fluctuations.}

At the order retained in the high-temperature expansion, the
thermally extensive part of the Hubbard--Stratonovich fluctuation
determinant is local.  On the leading saddle,
\begin{equation}
	T_{\rm loc}(\theta)
	=
	T\Gamma(\theta),
	\qquad
	\frac{\sqrt{\sigma(\theta)}}
	{T_{\rm loc}(\theta)}
	=
	x.
	\label{app:local-fluctuation-scales}
\end{equation}
After local frequencies and momenta are measured in units of
$T_{\rm loc}$, the finite thermal part contains no additional local
dimensionful scale.  Its thermally extensive contribution therefore
has the form
\begin{equation}
	\Delta_{\rm loc}\log Z
	=
	\beta
	\int_{S^2}dA\,
	T_{\rm loc}(\theta)^3\,
	f_{\rm loc}(x,\{\rho_n^*\}),
	\label{app:local-fluctuation-pressure}
\end{equation}
where $f_{\rm loc}$ is a dimensionless function of the saddle data.  We
do not need its explicit form; for the present scaling argument we only
require that it remain finite for finite saddle data.

The angular dependence then comes from the local temperature, and
hence
\begin{align}
	\Delta_{\rm loc}\log Z
	&=
	O\left(
	T^2
	\int_{S^2}dA\,\Gamma(\theta)^3
	\right)
	\nonumber\\
	&=
	O\left(
	\frac{(Tr)^2}{1-\omega^2}
	\right),
	\label{app:local-fluctuation-scaling}
\end{align}
where we used
\begin{equation}
	\int_{S^2}dA\,\Gamma(\theta)^3
	=
	2\pi r^2
	\int_0^\pi
	\frac{\sin\theta\,d\theta}
	{\left(1-\omega^2\sin^2\theta\right)^{3/2}}
	=
	\frac{4\pi r^2}{1-\omega^2}.
	\label{app:sigma-phase-space-integral}
\end{equation}
Thus the thermally extensive local fluctuation determinant has at most
a simple fast-rotation pole.

\paragraph{Correction induced by the holonomy saddle.}

The contribution $K_\rho$ in
equation~\eqref{app:rho-induced-kernel} is bilocal rather than a second
local thermal determinant.  Its structure becomes particularly simple
on the leading saddle.  From
equation~\eqref{crit-winding-coefficient},
\begin{equation}
	\frac{\delta A_n}{\delta\sigma(\theta)}
	=
	-\frac{r^2}{2}
	\sin\theta\,\Gamma(\theta)
	\exp\left[
	-\frac{n\sqrt{\sigma(\theta)}}
	{T\Gamma(\theta)}
	\right].
	\label{app:An-first-variation}
\end{equation}
Using
\begin{equation}
	\frac{\sqrt{\sigma(\theta)}}{T\Gamma(\theta)}=x,
\end{equation}
one obtains
\begin{equation}
	\frac{\delta A_n}{\delta\sigma(\theta)}
	=
	-\frac{r^2}{2}
	e^{-nx}
	\sin\theta\,\Gamma(\theta).
	\label{app:An-first-variation-onshell}
\end{equation}
Consequently,
\begin{align}
	K_\rho(\theta,\theta')
	&=
	C_\rho(x)\,
	f(\theta)f(\theta'),
	\label{app:rho-rank-one}
	\\
	f(\theta)
	&=
	\sin\theta\,\Gamma(\theta),
	\nonumber\\
	C_\rho(x)
	&=
	\frac{r^4}{4}
	\sum_{n=1}^{\infty}\frac{e^{-2nx}}{n}
	=
	-\frac{r^4}{4}
	\log\left(1-e^{-2x}\right).
	\label{app:rho-coefficient}
\end{align}
For $x>0$, and in particular on the transition saddle $x=x_c$, the
coefficient $C_\rho(x)$ is finite.

Equation~\eqref{app:rho-rank-one} shows that the feedback of the
holonomy saddle is a rank-one bilocal deformation of the local
Hubbard--Stratonovich kernel.  It therefore does not introduce a second
independent local phase-space trace.  We regard these kernels as integral
operators on $0\leq\theta\leq\pi$ with measure $d\theta$.  The inverse kernel
is defined by $\int_0^\pi d\vartheta\,
K_{\rm loc}(\theta,\vartheta)K_{\rm loc}^{-1}(\vartheta,\theta')
=\delta(\theta-\theta')$.  The rank-one determinant lemma then gives
\begin{equation}
	\det K_{\rm eff}
	=
	\det K_{\rm loc}
	\left[
	1+
	C_\rho(x)
	\int_0^\pi d\theta\int_0^\pi d\theta'\,
	f(\theta)K_{\rm loc}^{-1}(\theta,\theta')f(\theta')
	\right].
	\label{app:determinant-lemma}
\end{equation}
Hence the holonomy-induced part of the Gaussian correction takes the
form
\begin{equation}
	\Delta_\rho\log Z
	=
	-\frac12
	\log\left[
	1+
	C_\rho(x)
	\int_0^\pi d\theta\int_0^\pi d\theta'\,
	f(\theta)K_{\rm loc}^{-1}(\theta,\theta')f(\theta')
	\right].
	\label{app:rho-fluctuation-determinant}
\end{equation}
Provided no additional zero mode develops in the
Hubbard--Stratonovich kernel, any power-law enhancement of the quantity
inside the logarithm can produce at most a logarithmic fast-rotation
singularity.  In particular, motion of the $\rho_n$ saddle does not
generate an additional local phase-space factor and therefore cannot
produce the double pole found in the saddle contribution.

In the ordered strict-$N$ limit discussed in the main text one has
$x\to0$.  In this limit,
\begin{equation}
	C_\rho(x)
	=
	-\frac{r^4}{4}
	\log(2x)
	+O(x),
	\qquad
	x\to0^+ .
	\label{app:rho-small-x}
\end{equation}
Thus the coefficient itself develops only a logarithmic dependence on
$x$.  This does not supply an additional power of
$(1-\omega^2)^{-1}$ and therefore does not alter the power-law
fast-rotation comparison made below.

Combining the local and holonomy-induced pieces, the Gaussian
correction is therefore parametrically of at most simple-pole order,
\begin{equation}
	\Delta_{\rm fluc}\log Z
	=
	O\left(
	\frac{(Tr)^2}{1-\omega^2}
	\right)
	+
	O\left(
	\log\frac{1}{1-\omega^2}
	\right),
	\label{app:total-fluctuation-scaling}
\end{equation}
up to logarithmic dependence on $x$ and finite dimensionless
coefficients.

\paragraph{Collective-field measure.}

There is in addition a contribution from the collective-field measure
associated with replacing the finite set of eigenangles by the
continuum density $\rho(\alpha)$.  As discussed in
section~\ref{sec:free-density}, its logarithm is of order $N$ for a
smooth density.  It is therefore subleading to the $O(N^2)$ saddle on
the Gross--Witten--Wadia transition curve and cannot modify the leading
transition equations.

Its possible effect on the strict-$N$ fast-rotation behavior can also
be estimated parametrically.  Write its contribution to the exponent
schematically as
\begin{equation}
	{\cal W}_{\rm coll}
	=
	N\,{\cal S}_{\rm coll}[\rho]
	+O(N^0).
	\label{app:collective-measure}
\end{equation}
In the ungapped phase, compare the matter-dependent saddle with the
matter-free saddle.  Their difference is
\begin{equation}
	\delta\rho_n
	=
	\frac{A_n}{N}.
	\label{app:collective-delta-rho}
\end{equation}
Assuming that ${\cal S}_{\rm coll}[\rho]$ is regular about the uniform
ungapped saddle, its matter-dependent variation has the schematic
expansion
\begin{equation}
	N\,\delta{\cal S}_{\rm coll}
	=
	O(N\delta\rho)
	+
	O(N\delta\rho^2)
	+\cdots .
	\label{app:collective-expansion}
\end{equation}
Using equation~\eqref{app:collective-delta-rho},
\begin{equation}
	N\delta\rho_n
	=
	O(A_n),
	\qquad
	N(\delta\rho_n)^2
	=
	O\left(\frac{A_n^2}{N}\right).
	\label{app:collective-scaling}
\end{equation}
In the ordered strict-$N$ limit,
\begin{equation}
	A_n
	=
	O\left(
	\frac{(Tr)^2}{1-\omega^2}
	\right)
\end{equation}
for each fixed $n$.  Therefore the linear matter-dependent correction
from the collective measure is at most of simple-pole order, while the
quadratic and higher terms vanish after $N\to\infty$ at fixed
$Tr$ and fixed $|\omega|<1$.  The collective-field measure can thus
modify subleading terms but cannot generate the double pole
proportional to $A_n^2$.

\paragraph{Comparison with the saddle contribution.}

On the Gross--Witten--Wadia transition curve,
\begin{equation}
	\frac{(Tr)^2}{1-\omega^2}
	=
	O(N),
	\qquad
	x=x_c=O(N^0).
	\label{app:transition-fluctuation-scaling}
\end{equation}
The local Gaussian correction in
equation~\eqref{app:total-fluctuation-scaling}, as well as the
collective-field contribution, is consequently at most $O(N)$,
whereas the first and third terms of
equation~\eqref{crit-logZ-onshell} are $O(N^2)$.  The
holonomy-induced finite-rank part is still less singular.
Consequently, fluctuations and collective-measure effects cannot
modify the leading large-$N$ transition equations or the numerical
critical values obtained from them.

A different comparison applies in the ordered strict-$N$ limit used to
study the fast-rotation behavior of the ungapped phase.  The leading
saddle term behaves as
\begin{equation}
	\frac12
	\sum_{n=1}^{\infty}\frac{A_n^2}{n}
	=
	O\left(
	\frac{(Tr)^4}{(1-\omega^2)^2}
	\right),
	\label{app:sigma-saddle-comparison}
\end{equation}
whereas the thermally extensive Gaussian correction and the
matter-dependent collective-measure contribution are at most of
simple-pole order.  The correction induced by motion of the $\rho_n$
saddle is finite rank and is still less singular in the fast-rotation
power counting.  These effects may therefore modify the coefficient
of the simple-pole, logarithmic and finite terms, but they cannot
modify the double-pole coefficient obtained from the saddle.

We emphasize that this appendix establishes only the parametric order
of the omitted corrections; it does not compute their complete finite
contribution.  Moreover, precisely at the
Gross--Witten--Wadia endpoint the positivity constraint on the
eigenvalue density becomes active, so the detailed subleading
fluctuation theory need not remain purely Gaussian.  This does not
affect the leading $O(N^2)$ transition equations discussed in the main
text.

\bibliographystyle{JHEP}
\bibliography{references}

\providecommand{\href}[2]{#2}\begingroup\raggedright\begin{thebibliography}{10}

\bibitem{Klebanov:2002ja}
I.~R. Klebanov and A.~M. Polyakov, {\it {AdS dual of the critical O(N) vector
  model}},  {\em Phys. Lett. B} {\bf 550} (2002) 213--219,
  [\href{http://arxiv.org/abs/hep-th/0210114}{{\tt hep-th/0210114}}].

\bibitem{Leonhardt:2002sn}
T.~Leonhardt, A.~Meziane, and W.~Ruhl, {\it {On the proposed AdS dual of the
  critical O(N) sigma model for any dimension 2 {\ensuremath{<}} d
  {\ensuremath{<}} 4}},  {\em Phys. Lett. B} {\bf 555} (2003) 271--278,
  [\href{http://arxiv.org/abs/hep-th/0211092}{{\tt hep-th/0211092}}].

\bibitem{Petkou:2003zz}
A.~C. Petkou, {\it {Evaluating the AdS dual of the critical O(N) vector
  model}},  {\em JHEP} {\bf 03} (2003) 049,
  [\href{http://arxiv.org/abs/hep-th/0302063}{{\tt hep-th/0302063}}].

\bibitem{Das:2003vw}
S.~R. Das and A.~Jevicki, {\it {Large N collective fields and holography}},
  {\em Phys. Rev. D} {\bf 68} (2003) 044011,
  [\href{http://arxiv.org/abs/hep-th/0304093}{{\tt hep-th/0304093}}].

\bibitem{Sezgin:2003pt}
E.~Sezgin and P.~Sundell, {\it {Holography in 4D (super) higher spin theories
  and a test via cubic scalar couplings}},  {\em JHEP} {\bf 07} (2005) 044,
  [\href{http://arxiv.org/abs/hep-th/0305040}{{\tt hep-th/0305040}}].

\bibitem{Giombi:2009wh}
S.~Giombi and X.~Yin, {\it {Higher Spin Gauge Theory and Holography: The
  Three-Point Functions}},  {\em JHEP} {\bf 09} (2010) 115,
  [\href{http://arxiv.org/abs/0912.3462}{{\tt arXiv:0912.3462}}].

\bibitem{deMelloKoch:2010wdf}
R.~de~Mello~Koch, A.~Jevicki, K.~Jin, and J.~P. Rodrigues, {\it {$AdS_4/CFT_3$
  Construction from Collective Fields}},  {\em Phys. Rev. D} {\bf 83} (2011)
  025006, [\href{http://arxiv.org/abs/1008.0633}{{\tt arXiv:1008.0633}}].

\bibitem{Giombi:2011ya}
S.~Giombi and X.~Yin, {\it {On Higher Spin Gauge Theory and the Critical O(N)
  Model}},  {\em Phys. Rev. D} {\bf 85} (2012) 086005,
  [\href{http://arxiv.org/abs/1105.4011}{{\tt arXiv:1105.4011}}].

\bibitem{Giombi:2012ms}
S.~Giombi and X.~Yin, {\it {The Higher Spin/Vector Model Duality}},  {\em J.
  Phys. A} {\bf 46} (2013) 214003, [\href{http://arxiv.org/abs/1208.4036}{{\tt
  arXiv:1208.4036}}].

\bibitem{Giombi:2013fka}
S.~Giombi and I.~R. Klebanov, {\it {One Loop Tests of Higher Spin AdS/CFT}},
  {\em JHEP} {\bf 12} (2013) 068, [\href{http://arxiv.org/abs/1308.2337}{{\tt
  arXiv:1308.2337}}].

\bibitem{deMelloKoch:2014vnt}
R.~de~Mello~Koch, A.~Jevicki, J.~P. Rodrigues, and J.~Yoon, {\it {Canonical
  Formulation of $O(N)$ Vector/Higher Spin Correspondence}},  {\em J. Phys. A}
  {\bf 48} (2015), no.~10 105403, [\href{http://arxiv.org/abs/1408.4800}{{\tt
  arXiv:1408.4800}}].

\bibitem{deMelloKoch:2014mos}
R.~de~Mello~Koch, A.~Jevicki, J.~P. Rodrigues, and J.~Yoon, {\it {Holography as
  a Gauge Phenomenon in Higher Spin Duality}},  {\em JHEP} {\bf 01} (2015) 055,
  [\href{http://arxiv.org/abs/1408.1255}{{\tt arXiv:1408.1255}}].

\bibitem{deMelloKoch:2018ivk}
R.~de~Mello~Koch, A.~Jevicki, K.~Suzuki, and J.~Yoon, {\it {AdS Maps and
  Diagrams of Bi-local Holography}},  {\em JHEP} {\bf 03} (2019) 133,
  [\href{http://arxiv.org/abs/1810.02332}{{\tt arXiv:1810.02332}}].

\bibitem{Aharony:2020omh}
O.~Aharony, S.~M. Chester, and E.~Y. Urbach, {\it {A Derivation of AdS/CFT for
  Vector Models}},  {\em JHEP} {\bf 03} (2021) 208,
  [\href{http://arxiv.org/abs/2011.06328}{{\tt arXiv:2011.06328}}].

\bibitem{Sundborg:1999ue}
B.~Sundborg, {\it {The Hagedorn Transition, Deconfinement and $\mathcal{N}=4$
  SYM Theory}},  {\em Nucl. Phys. B} {\bf 573} (2000) 349--363,
  [\href{http://arxiv.org/abs/hep-th/9908001}{{\tt hep-th/9908001}}].

\bibitem{Aharony:2003sx}
O.~Aharony, J.~Marsano, S.~Minwalla, K.~Papadodimas, and M.~Van~Raamsdonk, {\it
  {The Hagedorn/Deconfinement Phase Transition in Weakly Coupled Large $N$
  Gauge Theories}},  {\em Adv. Theor. Math. Phys.} {\bf 8} (2004) 603--696,
  [\href{http://arxiv.org/abs/hep-th/0310285}{{\tt hep-th/0310285}}].

\bibitem{Jain:2013py}
S.~Jain, S.~Minwalla, T.~Sharma, T.~Takimi, S.~R. Wadia, and S.~Yokoyama, {\it
  {Phases of Large $N$ Vector Chern--Simons Theories on $S^2\times S^1$}},
  {\em JHEP} {\bf 09} (2013) 009, [\href{http://arxiv.org/abs/1301.6169}{{\tt
  arXiv:1301.6169}}].

\bibitem{Iliesiu:2018fao}
L.~Iliesiu, M.~Kolo\u{g}lu, R.~Mahajan, E.~Perlmutter, and D.~Simmons-Duffin,
  {\it {The Conformal Bootstrap at Finite Temperature}},  {\em JHEP} {\bf 10}
  (2018) 070, [\href{http://arxiv.org/abs/1802.10266}{{\tt arXiv:1802.10266}}].

\bibitem{Petkou:2018ynm}
A.~C. Petkou and A.~Stergiou, {\it {Dynamics of Finite-Temperature Conformal
  Field Theories from Operator Product Expansion Inversion Formulas}},  {\em
  Phys. Rev. Lett.} {\bf 121} (2018), no.~7 071602,
  [\href{http://arxiv.org/abs/1806.02340}{{\tt arXiv:1806.02340}}].

\bibitem{David:2023uya}
J.~R. David and S.~Kumar, {\it {Thermal one-point functions:
  CFT{\textquoteright}s with fermions, large d and large spin}},  {\em JHEP}
  {\bf 10} (2023) 143, [\href{http://arxiv.org/abs/2307.14847}{{\tt
  arXiv:2307.14847}}].

\bibitem{David:2024naf}
J.~R. David and S.~Kumar, {\it {One point functions in large N vector models at
  finite chemical potential}},  {\em JHEP} {\bf 01} (2025) 080,
  [\href{http://arxiv.org/abs/2406.14490}{{\tt arXiv:2406.14490}}].

\bibitem{Kumar:2025txh}
S.~Kumar, {\it {Large N Wess-Zumino model at finite temperature and large
  chemical potential in 3d}},  {\em JHEP} {\bf 08} (2025) 148,
  [\href{http://arxiv.org/abs/2503.17999}{{\tt arXiv:2503.17999}}].

\bibitem{Benedetti:2023pbt}
D.~Benedetti, R.~Gurau, S.~Harribey, and D.~Lettera, {\it {Finite-size versus
  finite-temperature effects in the critical long-range O(N) model}},  {\em
  JHEP} {\bf 02} (2024) 078, [\href{http://arxiv.org/abs/2311.04607}{{\tt
  arXiv:2311.04607}}].

\bibitem{Diatlyk:2023msc}
O.~Diatlyk, F.~K. Popov, and Y.~Wang, {\it {Beyond~$N=\infty$ in Large $N$
  conformal vector models at finite temperature}},  {\em JHEP} {\bf 08} (2024)
  219, [\href{http://arxiv.org/abs/2309.02347}{{\tt arXiv:2309.02347}}].

\bibitem{Barrat:2025wbi}
J.~Barrat, E.~Marchetto, A.~Miscioscia, and E.~Pomoni, {\it {Thermal Bootstrap
  for the Critical O(N) Model}},  {\em Phys. Rev. Lett.} {\bf 134} (2025),
  no.~21 211604, [\href{http://arxiv.org/abs/2411.00978}{{\tt
  arXiv:2411.00978}}].

\bibitem{Marchetto:2023fcw}
E.~Marchetto, A.~Miscioscia, and E.~Pomoni, {\it {Broken (super) conformal Ward
  identities at finite temperature}},  {\em JHEP} {\bf 12} (2023) 186,
  [\href{http://arxiv.org/abs/2306.12417}{{\tt arXiv:2306.12417}}].

\bibitem{Marchetto:2023xap}
E.~Marchetto, A.~Miscioscia, and E.~Pomoni, {\it {Sum rules {\&} Tauberian
  theorems at finite temperature}},  {\em JHEP} {\bf 09} (2024) 044,
  [\href{http://arxiv.org/abs/2312.13030}{{\tt arXiv:2312.13030}}].

\bibitem{Niarchos:2025cdg}
V.~Niarchos, C.~Papageorgakis, A.~Stratoudakis, and M.~Woolley, {\it {Deep
  Finite Temperature Bootstrap}},  \href{http://arxiv.org/abs/2508.08560}{{\tt
  arXiv:2508.08560}}.

\bibitem{Barrat:2025nvu}
J.~Barrat, D.~N. Bozkurt, E.~Marchetto, A.~Miscioscia, and E.~Pomoni, {\it {The
  analytic bootstrap at finite temperature}},  {\em JHEP} {\bf 05} (2026) 104,
  [\href{http://arxiv.org/abs/2506.06422}{{\tt arXiv:2506.06422}}].

\bibitem{David:2024pir}
J.~R. David and S.~Kumar, {\it {The large N vector model on S$^{1}$
  {\texttimes} S$^{2}$}},  {\em JHEP} {\bf 03} (2025) 169,
  [\href{http://arxiv.org/abs/2411.18509}{{\tt arXiv:2411.18509}}].

\bibitem{David:2025tqn}
J.~R. David and S.~Kumar, {\it {High to low temperature: O(N) model at large
  N}},  {\em JHEP} {\bf 02} (2026) 194,
  [\href{http://arxiv.org/abs/2508.14872}{{\tt arXiv:2508.14872}}].

\bibitem{Giombi:2014iua}
S.~Giombi, I.~R. Klebanov, and A.~A. Tseytlin, {\it {Partition Functions and
  Casimir Energies in Higher Spin AdS$_{d+1}$/CFT$_d$}},  {\em Phys. Rev. D}
  {\bf 90} (2014), no.~2 024048, [\href{http://arxiv.org/abs/1402.5396}{{\tt
  arXiv:1402.5396}}].

\bibitem{Gross:1980he}
D.~J. Gross and E.~Witten, {\it {Possible Third Order Phase Transition in the
  Large N Lattice Gauge Theory}},  {\em Phys. Rev. D} {\bf 21} (1980) 446--453.

\bibitem{Wadia:1980cp}
S.~R. Wadia, {\it {N = Infinity Phase Transition in a Class of Exactly Soluble
  Model Lattice Gauge Theories}},  {\em Phys. Lett. B} {\bf 93} (1980)
  403--410.

\bibitem{Shenker:2011zf}
S.~H. Shenker and X.~Yin, {\it {Vector Models in the Singlet Sector at Finite
  Temperature}},  \href{http://arxiv.org/abs/1109.3519}{{\tt arXiv:1109.3519}}.

\bibitem{Mauro:2026zus}
L.~Mauro and A.~Vichi, {\it {Thermal effective action for the $O(N)$ vector
  model}},  \href{http://arxiv.org/abs/2606.05059}{{\tt arXiv:2606.05059}}.

\bibitem{David:2026yis}
J.~R. David and S.~Kumar, {\it {The large $N$ vector model with angular
  velocity}},  \href{http://arxiv.org/abs/2608.31151}{{\tt arXiv:2608.31151}}.

\bibitem{Mishra:2026bpi}
V.~Mishra, {\it {Semi-Universality of $O(N)$ model on $S^1\times S^2$}},
  \href{http://arxiv.org/abs/2609.25622}{{\tt arXiv:2609.25622}}.

\bibitem{Anand:2025mfh}
H.~Anand, N.~Benjamin, V.~Kumar, S.~Minwalla, J.~Mukherjee, S.~Pal, and
  A.~Rahaman, {\it {Semi-universality of CFT$_d$ entropy at large spin}},  {\em
  SciPost Phys.} {\bf 21} (2026) 042,
  [\href{http://arxiv.org/abs/2512.00158}{{\tt arXiv:2512.00158}}].

\bibitem{Cunden_2018}
F.~D. Cunden, F.~Mezzadri, and N.~O’Connell, {\it Free fermions and the
  classical compact groups},  {\em Journal of Statistical Physics} {\bf 171}
  (Apr., 2018) 768–801.

\bibitem{Meckes_2019}
E.~S. Meckes, {\em The Random Matrix Theory of the Classical Compact Groups}.
\newblock Cambridge Tracts in Mathematics. Cambridge University Press, 2019.

\bibitem{JevickiSakita1980}
A.~Jevicki and B.~Sakita, {\it {The Quantum Collective Field Method and Its
  Application to the Planar Limit}},  {\em Nucl. Phys. B} {\bf 165} (1980)
  511--527.

\bibitem{DeanMajumdar2008}
D.~S. Dean and S.~N. Majumdar, {\it {Extreme Value Statistics of Eigenvalues of
  Gaussian Random Matrices}},  {\em Phys. Rev. E} {\bf 77} (2008) 041108,
  [\href{http://arxiv.org/abs/0801.1730}{{\tt arXiv:0801.1730}}].

\bibitem{Maldacena:2011jn}
J.~Maldacena and A.~Zhiboedov, {\it {Constraining Conformal Field Theories with
  A Higher Spin Symmetry}},  {\em J. Phys. A} {\bf 46} (2013) 214011,
  [\href{http://arxiv.org/abs/1112.1016}{{\tt arXiv:1112.1016}}].

\bibitem{Maldacena:2012sf}
J.~Maldacena and A.~Zhiboedov, {\it {Constraining conformal field theories with
  a slightly broken higher spin symmetry}},  {\em Class. Quant. Grav.} {\bf 30}
  (2013) 104003, [\href{http://arxiv.org/abs/1204.3882}{{\tt
  arXiv:1204.3882}}].

\bibitem{Alba:2013yda}
V.~Alba and K.~Diab, {\it {Constraining conformal field theories with a higher
  spin symmetry in d=4}},  \href{http://arxiv.org/abs/1307.8092}{{\tt
  arXiv:1307.8092}}.

\bibitem{Alba:2015upa}
V.~Alba and K.~Diab, {\it {Constraining conformal field theories with a higher
  spin symmetry in $d > 3$ dimensions}},  {\em JHEP} {\bf 03} (2016) 044,
  [\href{http://arxiv.org/abs/1510.02535}{{\tt arXiv:1510.02535}}].

\bibitem{Advant:2026cqt}
A.~Advant, H.~Anand, N.~Benjamin, V.~Kumar, S.~Minwalla, J.~Mukherjee, S.~Pal,
  A.~Rahaman, and P.~Ray, {\it {From $Z$ to $a$: High-temperature relations,
  subleading semi-universality, and conformal anomalies}},
  \href{http://arxiv.org/abs/2609.03013}{{\tt arXiv:2609.03013}}.

\bibitem{Cassani:2021fyv}
D.~Cassani and Z.~Komargodski, {\it {EFT and the SUSY Index on the 2nd Sheet}},
   {\em SciPost Phys.} {\bf 11} (2021) 004,
  [\href{http://arxiv.org/abs/2104.01464}{{\tt arXiv:2104.01464}}].

\bibitem{Benjamin:2023qsc}
N.~Benjamin, J.~Lee, H.~Ooguri, and D.~Simmons-Duffin, {\it {Universal
  asymptotics for high energy CFT data}},  {\em JHEP} {\bf 03} (2024) 115,
  [\href{http://arxiv.org/abs/2306.08031}{{\tt arXiv:2306.08031}}].

\bibitem{Mukherjee:2026dfu}
J.~Mukherjee and P.~Ray, {\it {Semi-universality of conformal higher-derivative
  and conformal higher-spin fields}},  {\em JHEP} {\bf 08} (2026) 182,
  [\href{http://arxiv.org/abs/2606.08461}{{\tt arXiv:2606.08461}}].

\bibitem{Komargodski:2026ain}
Z.~Komargodski, A.~Miscioscia, and F.~K. Popov, {\it {Regge's Inferno}},
  \href{http://arxiv.org/abs/2603.10197}{{\tt arXiv:2603.10197}}.

\bibitem{Amado:2016pgy}
I.~Amado, B.~Sundborg, L.~Thorlacius, and N.~Wintergerst, {\it {Probing
  emergent geometry through phase transitions in free vector and matrix
  models}},  {\em JHEP} {\bf 02} (2017) 005,
  [\href{http://arxiv.org/abs/1612.03009}{{\tt arXiv:1612.03009}}].

\bibitem{Romatschke:2019ybu}
P.~Romatschke, {\it {Finite-Temperature Conformal Field Theory Results for All
  Couplings: O(N) Model in 2+1 Dimensions}},  {\em Phys. Rev. Lett.} {\bf 122}
  (2019), no.~23 231603, [\href{http://arxiv.org/abs/1904.09995}{{\tt
  arXiv:1904.09995}}]. [Erratum: Phys.Rev.Lett. 123, 209901 (2019)].

\bibitem{Grable:2022swa}
S.~Grable, {\it {Interacting CFTs for all couplings: thermal versus
  entanglement entropy at large N}},  {\em JHEP} {\bf 10} (2022) 133,
  [\href{http://arxiv.org/abs/2205.15383}{{\tt arXiv:2205.15383}}].

\bibitem{Giombi:2011kc}
S.~Giombi, S.~Minwalla, S.~Prakash, S.~P. Trivedi, S.~R. Wadia, and X.~Yin,
  {\it {Chern-Simons Theory with Vector Fermion Matter}},  {\em Eur. Phys. J.
  C} {\bf 72} (2012) 2112, [\href{http://arxiv.org/abs/1110.4386}{{\tt
  arXiv:1110.4386}}].

\bibitem{Aharony:2011jz}
O.~Aharony, G.~Gur-Ari, and R.~Yacoby, {\it {d=3 Bosonic Vector Models Coupled
  to Chern-Simons Gauge Theories}},  {\em JHEP} {\bf 03} (2012) 037,
  [\href{http://arxiv.org/abs/1110.4382}{{\tt arXiv:1110.4382}}].

\bibitem{Parmentier:2026aqh}
K.~Parmentier and N.~Bobev, {\it {CFTs on Squashed Spheres and the Thermal
  Effective Action}},  \href{http://arxiv.org/abs/2606.30716}{{\tt
  arXiv:2606.30716}}.

\bibitem{Lawrence:2023woz}
S.~Lawrence, R.~Weller, C.~Peterson, and P.~Romatschke, {\it {Instantons,
  analytic continuation, and PT-symmetric field theory}},  {\em Phys. Rev. D}
  {\bf 108} (2023), no.~8 085013, [\href{http://arxiv.org/abs/2303.01470}{{\tt
  arXiv:2303.01470}}].

\bibitem{Grable:2023paf}
S.~Grable and M.~Weiner, {\it {A fully solvable model of fermionic interaction
  in 3 + 1d}},  {\em JHEP} {\bf 09} (2023) 017,
  [\href{http://arxiv.org/abs/2302.08603}{{\tt arXiv:2302.08603}}].

\bibitem{Buric:2024kxo}
I.~Buric, F.~Russo, V.~Schomerus, and A.~Vichi, {\it {Thermal one-point
  functions and their partial wave decomposition}},  {\em JHEP} {\bf 12} (2024)
  021, [\href{http://arxiv.org/abs/2408.02747}{{\tt arXiv:2408.02747}}].

\bibitem{Buric:2025uqt}
I.~Buri{\'c}, F.~Mangialardi, F.~Russo, V.~Schomerus, and A.~Vichi, {\it
  {Heavy-heavy-light asymptotics from thermal correlators}},  {\em JHEP} {\bf
  04} (2026) 027, [\href{http://arxiv.org/abs/2506.21671}{{\tt
  arXiv:2506.21671}}].

\bibitem{Ammon:2025cdz}
M.~Ammon, J.~Hollweck, T.~H{\"o}ssel, and K.~W{\"o}lfl, {\it {Thermal n-point
  conformal blocks in four dimensions from oscillator representations}},  {\em
  JHEP} {\bf 04} (2026) 102, [\href{http://arxiv.org/abs/2507.22974}{{\tt
  arXiv:2507.22974}}].

\bibitem{Miscioscia:2026bvl}
A.~Miscioscia, D.~Pavshinkin, and F.~K. Popov, {\it {Regge in Wonderland}},
  \href{http://arxiv.org/abs/2609.35455}{{\tt arXiv:2609.35455}}.

\bibitem{NIST:DLMF}
``{\it NIST Digital Library of Mathematical Functions}.''
  \url{https://dlmf.nist.gov/}, Release 1.2.7 of 2026-06-15.
\newblock F.~W.~J. Olver, A.~B. {Olde Daalhuis}, D.~W. Lozier, B.~I. Schneider,
  R.~F. Boisvert, C.~W. Clark, B.~R. Miller, B.~V. Saunders, H.~S. Cohl, and
  M.~A. McClain, eds.

\bibitem{DimassiSjostrand1999}
M.~Dimassi and J.~Sj{\"o}strand, {\em Spectral Asymptotics in the
  Semi-Classical Limit}, vol.~268 of {\em London Mathematical Society Lecture
  Note Series}.
\newblock Cambridge University Press, 1999.

\end{thebibliography}\endgroup

\end{document}